\documentclass[aps,showpacs,twocolumn,superscriptaddress]{revtex4}
\usepackage{amsmath}
\usepackage{amssymb}
\usepackage{amscd}
\usepackage{wasysym}
\usepackage[ansinew]{inputenc}
\usepackage[T1]{fontenc}
\usepackage{ae,aecompl}

\usepackage[dvips]{graphicx}
\DeclareGraphicsExtensions{.eps}
  
\newcommand{\be}{\begin{equation}}
\newcommand{\ee}{\end{equation}}
\newcommand{\bea}{\begin{eqnarray}}
\newcommand{\eea}{\end{eqnarray}}
\newcommand{\nn}{\nonumber}
\newcommand{\ket}[1]{|#1\rangle}
\newcommand{\bra}[1]{\left\langle#1\right|}

\newcommand{\eye}{\mbox{$\mbox{1}\!\mbox{l}\;$}}

\renewcommand{\vec}[1]{\mathbf{#1}}

\newcommand{\tr}{{\rm tr}}

\begin{document}
\bibliographystyle{apsrev}

\title{Dissipation induced coherence and stochastic resonance of an open two-mode Bose-Einstein condensate}
\author{D. Witthaut}
\email{dirk.witthaut@nbi.dk}
\affiliation{QUANTOP, Niels Bohr Institute, University of Copenhagen, 
DK--2100 Copenhagen, Denmark}
\author{F. Trimborn}
\affiliation{Institut f\"ur mathematische Physik, TU Braunschweig, D--38106 Braunschweig, Germany}
\author{S. Wimberger}
\affiliation{Institut f\"ur theoretische Physik, Universit\"{a}t Heidelberg, D--69120, Heidelberg, Germany}
\date{\today }

\begin{abstract}
We discuss the dynamics of a Bose-Einstein condensate in a 
double-well trap subject to phase noise and particle loss. 
The phase coherence of a weakly-interacting condensate, experimentally 
measured via the contrast in an interference experiment, as well as 
the response to an external driving become maximal for a finite value 
of the dissipation rate matching the intrinsic time scales of the system. 
This can be understood as a stochastic resonance of the many-particle system. 
Even stronger effects are observed when dissipation acts in 
concurrence with strong inter-particle interactions, restoring
the purity of the condensate almost completely and increasing
the phase coherence significantly. Our theoretical results are backed  by 
Monte Carlo simulations, which show a good qualitative agreement and provide 
a microscopic explanation for the observed stochastic resonance effect.
\end{abstract}

\maketitle

\section{Introduction}

Stochastic resonance (SR) is a strongly surprising, yet very general effect
in nonlinear dynamical systems. Against our naive understanding, the response
of a system to an external driving  can be facilitated if an appropriate amount of 
noise is added. In fact, the maximum of the response -- the stochastic resonance -- 
is found if the timescale of the noise matches an intrinsic time scale of the system.
The effect was first described for strongly damped classical systems such as 
the overdamped particle in a driven double well trap. 
In this case the noise is strong enough to induce the transition between the wells, 
whereas it is still weak enough not to randomize the 
dynamics completely. The particle will then hop to and fro almost deterministically 
if the average transition time between the wells due to the noise equals half 
of the driving period \cite{Benz81}. By now, a stochastic resonance has been shown 
in a variety of systems, an overview can be found in the review articles 
\cite{Wies95,Dykm95,Gamm98,Well04}. In addition to numerous examples in classical 
dynamics, stochastic resonance has also been found in a variety of quantum systems 
(see, e.g., \cite{Lofs94a,Lofs94b,Buch98,Well99,Huel00,Adam01,Well04}).

Recently, there has been an increased interest in the effects of dissipation
and the possibilities to control these in interacting many-body quantum systems.
For instance, the entanglement in a spin chain assumes an SR-like maximum for 
a finite amount of thermal noise \cite{Huel07}. Methods to attenuate phase noise 
for an open two-mode BEC were discussed in \cite{Khod08}, and the effects of particle
loss on the spin squeezing of such a system were analyzed in \cite{Li08}.
Furthermore, it has been shown that dissipative processes can be tailored to 
prepare arbitrary pure states for quantum computation and strongly correlated
states of ultracold atoms \cite{Krau08,Diel08} or to implement a universal set of 
quantum gates \cite{Vers08}. Actually, a recent experiment has even proven 
that strong inelastic collisions may inhibit particle losses and induce 
strong correlations in a quasi one-dimensional gas of ultracold atoms
\cite{Syas08,Garc08}.

In the present paper, we investigate the effects of noise and dissipation for 
a Bose-Einstein condensate (BEC) in a double-well trap. The essential idea has 
been introduced in a recent letter \cite{08stores}, and here we extend the 
discussion to a detailed analysis of the predicted SR-phenomenon. The setup 
under consideration has been experimentally realized by several groups only 
in the last few years \cite{Albi05,Gati06a,Gati06b,Schu05b,Foll07,Trot08}. 
Ultracold atoms in optical potentials have the enormous advantage that they 
allow to observe the quantum dynamics of an interacting many-particle system 
\textit{in situ}. Thus they serve as excellent model systems, bringing together
aspects of nonlinear dynamics, solid-state physics and the theory of open
quantum systems.
Here we show that the coherence of the two condensate modes assumes a maximum
in the fashion of the stochastic resonance effect for a finite dissipation rate, 
which matches 
the time scales of the intrinsic dynamics. In this case the particle loss is 
strong enough to significantly increase the condensate purity, whereas it is 
still weak enough not to dominate the dynamics completely.
Similarly the response to an external driving is increased if a proper amount
of dissipation is present. Even more remarkable results are found when dissipation
acts in concurrence with strong inter-particle interactions, leading to an almost
complete revival of the purity of the BEC. These effects are of considerable 
strength for realistic parameters and thus should be readily observable in ongoing 
experiments.

This paper is organized as follows:
First, we introduce the theoretical description of the open two-mode Bose-Hubbard system. 
We discuss the main sources of noise and dissipation and derive the corresponding mean-field 
approximation of the many-particle system.
The resulting dynamics for weak inter-particle interactions is analyzed in 
Sec.~\ref{sec-dic-weak}. It is shown that the phase contrast between the two 
modes assumes an SR-like maximum if the time scales of tunneling and dissipation 
are matched. This result is explained within the mean-field approximation as well 
as for the underlying many-particle quantum dynamics with Monte Carlo simulations 
backing up the approximative results.
The response of the open system to an external driving is discussed in 
Sec.~\ref{sec-sr-weak}. The amplitude of the forced oscillation also shows
a pronounced stochastic resonance effect.
Sec.~\ref{sec-dic-strong} then investigates the case of a strongly interacting
BEC, which is a problem of both, fundamental theoretical interest as well as high 
experimental relevance.
The interplay between interactions and dissipation can restore the purity
of the condensate almost completely and significantly increase the phase 
coherence in comparison with situations where one of the two is weak or missing.
This counter-intuitive effect is robust and can be explained by the appearance 
of novel nonlinear eigenstates.

\section{Noise and dissipation in a trapped BEC}
\label{sec-noise}

\begin{figure}[tb]
\centering
\includegraphics[width=5cm, angle=0]{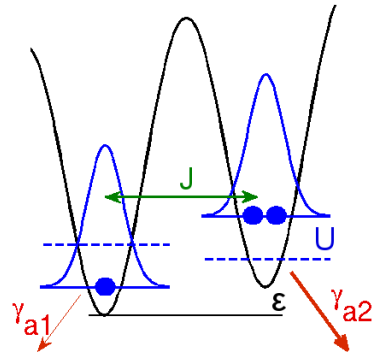}
\caption{\label{fig-bhmodel}
(Color online)
The open double-well trap considered in the present paper.}
\end{figure}

The basic setup under consideration is depicted in Fig.~\ref{fig-bhmodel}. 
Ultracold atoms are confined in a double-well trap that can be realized e.g. 
by superimposing an optical lattice with an optical dipole trap 
\cite{Albi05,Gati06a,Gati06b}, in a bichromatic optical lattice \cite{Foll07,Trot08}, 
or on an atom chip \cite{Schu05b}.
We consider the case where only one mode in each trap is significantly 
populated, whereas all higher modes contribute to the heat bath (see below). 
The unitary dynamics of the atoms is then described by the two-mode 
Bose-Hubbard Hamiltonian \cite{Milb97,Smer97,Vard01,Angl01}
\bea
  \hat H &=& - J  \left( \hat a_1^\dagger \hat a_2 +  \hat a_2^\dagger \hat a_1 \right)
     + \epsilon_2 \hat n_2 + \epsilon_1 \hat n_1   \nn \\
    && \qquad + \frac{U}{2} \left( \hat n_1(\hat n_1 - 1) + \hat n_2(\hat n_2 - 1)
      \right),
    \label{eqn-hami-bh}
\eea
where $\hat a_j$ and $\hat a_j^\dagger$ are the bosonic annihilation and 
creation operators in mode $j$ and $\hat n_j = \hat a_j^\dagger \hat a_j$
is the corresponding number operator. Furthermore, $J$ denotes the tunneling
matrix element between the wells, $U$ the interaction strength and $\epsilon_j$
the on-site energy of the $j$th well. We set $\hbar = 1$, thus measuring 
all energies in frequency units.

In order to clarify the algebraic structure of the model and to analyze 
the dynamics in the Bloch representation we introduce the collective operators
\bea
  \hat L_x &=& \frac{1}{2} 
      \left( \hat a_1^\dagger \hat a_2  + \hat a_2^\dagger \hat a_1 \right) \nn \\
  \hat L_y &=& \frac{i}{2} 
      \left( \hat a_1^\dagger \hat a_2  - \hat a_2^\dagger \hat a_1 \right) 
      \label{eqn-angular-op} \\
  \hat L_z &=& \frac{1}{2} 
      \left( \hat a_2^\dagger \hat a_2  - \hat a_1^\dagger \hat a_1 \right), \nn   
\eea
which form an angular momentum algebra $su(2)$ with quantum number $\ell = N/2$
\cite{Milb97,Smer97,Vard01,Angl01,08phase1,08phase2}, where $N$ is the actual
particle number. The Hamiltonian 
(\ref{eqn-hami-bh}) then can be rewritten  as
\be
  \hat H =  -2 J \hat L_x + 2\epsilon \hat L_z   + U \hat L_z^2
  \label{eqn-hamiltonian-2level}
\ee
up to terms only depending on the total number of atoms. Here, 
$\epsilon = \epsilon_2 - \epsilon_1$ denotes the difference of
the on-site energies of the two wells.

A model for noise and dissipation in a deep trapping potential has been
derived by Anglin \cite{Angl97} and later extended by Ruostekoski and Walls 
\cite{Ruos98} to the case of two weakly coupled modes. The dynamics is 
then given by the master equation
\bea
  \dot{\hat \rho} &=&   -i [\hat H,\hat \rho] 
  - \frac{\gamma_p}{2} \sum_{j = 1,2} \left(
  \hat n_j^2 \hat \rho + \hat \rho \hat n_j^2 - 2 \hat n_j \hat \rho \hat n_j \right)
    \label{eqn-master1} \\
  && - \frac{\gamma_a}{2}  \sum_{j=1,2;\pm} \left( \hat C_{j \pm}^\dagger \hat C_{j \pm} \hat  \rho
    + \hat \rho \hat C_{j \pm}^\dagger \hat C_{j \pm} 
    - 2 \hat C_{j \pm} \hat \rho \hat C_{j \pm}^\dagger \right)\nn
\eea
with the Lindblad operators
\bea
  \hat C_{j+} &=& \hat a_j^{\dagger} \qquad \mbox{and} \nn \\
  \hat C_{j-} &=&  e^{\beta/2(\epsilon_j - \mu + U \hat n_j)} \hat a_j ,
\eea
describing growth and depletion of the condensate.

Let us briefly discuss the effects of the noise and dissipation terms.
The second term $\sim \gamma_p$ in Eqn.~(\ref{eqn-master1}) describes 
phase noise due to elastic collisions with the background gas atoms.
It is usually the dominating contribution, effectively heating the 
system, but leaving the total particle number invariant. If only phase 
noise is present, the system relaxes to an equilibrium state where 
all coherences are lost and all Dicke states 
$\ket{n_1,N-n_1} \sim \hat a_1^{\dagger n_1}\hat a_2^{\dagger N-n_1}\ket{0,0}$ 
are equally populated 
\be
   \langle  n_1, N - n_1  | \hat \rho | n'_1, N- n'_1 \rangle
    = \frac{1}{N+1} \delta_{n_1,n'_1},
\ee
as long as $J \neq 0$ \cite{Wall85,Gard04}. This corresponds to a thermal 
state of infinite temperature with $\langle \hat {\vec L} \rangle = \vec 0$.
The remaining terms $\sim \gamma_a$ in the master equation (\ref{eqn-master1})
describe amplitude noise, i.e.~the growth and depletion of the condensate due
to inelastic collisions with the background gas. In contrast to phase noise,
amplitude noise heats {\it and} cools the system. If both amplitude and 
phase noise are present, the system will relax to the proper thermal state
with a density operator 
$\hat \rho \propto \exp(-\beta (\hat H - \mu \hat n))$ \cite{Angl97}. 

In current experiments amplitude noise and dissipation is usually 
extremely weak in comparison to phase noise \cite{Ruos98},
if it is not introduced artificially as for example by forced 
evaporative cooling during the preparation of the BEC. 
For example, phase noise damps Josephson oscillations on a 
timescale of a few hundred milliseconds in the experiments, 
while less than $10 \%$ of the atoms are lost during a 
$30 \, {\rm s}$ experiment \cite{Albi05,Gati06a,Gati06b}. 
This is much too weak to produce the effects discussed in the 
present paper. In contrast, a strong and tunable source of dissipation 
can be implemented artificially by shining a resonant laser beam 
onto the trap, that removes atoms with the site-dependent rates 
$\gamma_{aj}$ from the two wells $j = 1,2$. In magnetic trapping 
potentials, this can also be achieved by a forced rf-transition to an 
untrapped magnetic substate \cite{Bloc99}. Non-trivial effects of 
dissipation such as the stochastic resonance discussed below require 
strongly biased loss rates, i.e. $\gamma_{a1} \neq \gamma_{a2}$. 
For a laser beam focused on one of the wells an asymmetry of
$f_a = (\gamma_{a2}$ -- $\gamma_{a1})/ (\gamma_{a2}$ + $\gamma_{a1}) = 0.5$
should be feasible.
The master equation description of both, phase noise and particle loss, 
is well established \cite{Gard04} and routinely used in the context of
photon fields. In the following we will thus consider the dynamics 
generated by the master equation
\bea
  \dot{\hat \rho} &=& -i [\hat H,\hat \rho] 
  - \frac{\gamma_p}{2} \sum_{j = 1,2}
    \left( \hat n_j^2 \hat \rho + \hat \rho \hat n_j^2 
          - 2 \hat n_j \hat \rho \hat n_j  \right) \nn \\
   && \quad - \frac{1}{2}  \sum_{j=1,2}  \gamma_{aj} \left(
     \hat a_j^\dagger \hat a_j \hat \rho + \hat \rho \hat a_j^\dagger \hat a_j  
    - 2 \hat a_j \hat \rho \hat a_j^\dagger \right).
   \label{eqn-master2}
\eea

The macroscopic dynamics of the atomic cloud is to a very good 
approximation \cite{Vard01,Angl01,08mfdecay} described by a 
mean-field approximation, considering only the expectation values 
$s_j(t) = 2 \, \tr(\hat L_j \hat \rho(t))$ of the angular momentum 
operators (\ref{eqn-angular-op}) and the particle number 
$n(t) = \tr((\hat n_1 + \hat n_2) \hat \rho(t))$. The evolution 
equations for the Bloch vector $\vec s = (s_x,s_y,s_z)$ 
are then calculated starting from the Master equation via 
$\dot s_j = \tr(\hat L_j \dot {\hat \rho})$ with the 
exact result (cf. \cite{08mfdecay})
\bea
  \dot s_x &=& -2 \epsilon s_y - U (s_y s_z + \Delta_{yz}) - T_2^{-1} s_x, \nn \\
  \dot s_y &=&  2 J s_z + 2\epsilon  s_x + U (s_x s_z + \Delta_{xz})  - T_2^{-1} s_y, \nn \\
  \dot s_z &=& - 2 J s_y - T_1^{-1} s_z - T_1^{-1} f_a n, \nn \\
  \dot n   &=& - T_1^{-1} n  - T_1^{-1} f_a s_z,
  \label{eqn-eom-bloch}
\eea
where we have defined the transversal and longitudinal damping times
by
\be
  T_1^{-1} = (\gamma_{a1} + \gamma_{a2})/2 \quad \mbox{and} \quad
  T_2^{-1} = \gamma_p +  T_1^{-1}.
  \label{eqn-rel-times}
\ee
These equations of motion resemble the celebrated Bloch equations in 
nuclear magnetic resonance \cite{Bloc46,Viol00}, with some subtle but 
nevertheless important differences. The longitudinal relaxation is now
associated with particle loss and, more important, the dynamics is
substantially altered by the $U$-dependent interaction term 
\cite{Milb97,Smer97,Albi05}.

The exact equations of motion (\ref{eqn-eom-bloch}) still include the covariances
\be
  \Delta_{jk} = \langle \hat L_j \hat L_k + \hat L_k \hat L_j  \rangle 
   - 2 \langle \hat L_j \rangle \langle \hat L_k \rangle.
\ee
The celebrated mean-field description is now obtained by approximating 
the second order moments by products of expectation values such that 
$\Delta_{jk}\approx0$ \cite{Milb97,Smer97,Vard01,Angl01}. This truncation is 
valid in the macroscopic limit of large particle numbers, since the 
covariances vanish as $1/n$ if the many particle quantum state is 
close to a pure BEC.

In the following, we will show that a finite amount of dissipation
induces a maximum of the coherence which can be understood as an 
stochastic resonance effect.
In this discussion we have to distinguish between two different kinds
of coherence, which will both be considered in the following. First of 
all we consider the phase coherence between the two wells, which is 
measured by the average \textit{contrast} in interference experiments 
as described in \cite{Albi05,Gati06a,Gati06b} and given by
\be
  \alpha(t) = \frac{2|\langle \hat a_1^\dagger \hat a_2\rangle|
            }{\langle \hat n_1 + \hat n_2 \rangle } 
         = \frac{\sqrt{s_x(t)^2+s_y(t)^2}}{n(t)} \, .
  \label{eqn-alpha-def}
\ee
Secondly, we will analyze how close the many-particle quantum state
is to a pure Bose-Einstein condensate. This property is quantified 
by the purity 
\be
   p = 2 \, \tr(\hat \rho_{\rm red}^2)-1
\ee
of the reduced single-particle density matrix \cite{Vard01,Angl01,Legg01,08phase2}
\be
  \hat \rho_{\rm red} = \frac{1}{N} \left( \begin{array}{c c}
   \langle \hat a_1^\dagger \hat a_1 \rangle & \langle \hat a_1^\dagger \hat a_2 \rangle \\
   \langle \hat a_2^\dagger \hat a_1 \rangle & \langle \hat a_2^\dagger \hat a_2 \rangle \\
   \end{array} \right).
   \label{eqn-spdm}
\ee   
One can easily show that the purity is related to the Bloch vector $\vec s$
by $p = |\vec s|^2/n^2$. A pure BEC, corresponding to a product state, is,
of course, characterized by $p=1$.

\section{Dissipation induced coherence in a weakly-interacting BEC}
\label{sec-dic-weak}

In this section, we show that a proper amount of dissipation
can indeed increase the phase coherence (\ref{eqn-alpha-def}) of a two-mode 
BEC similar to the stochastic resonance effect. For simplicity, we start
with the linear case $U=0$, where the mean-field equations of motion for 
the expectation values (\ref{eqn-eom-bloch}) are exact. The linear equations
resemble the Bloch equations for driven nuclear spins in the rotating wave 
approximation \cite{Bloc46}, which are known to show a pronounced stochastic
resonance effect \cite{Viol00}: The amplitude of forced oscillations of the 
spins given by $s_y$ assumes a maximum for a finite value of the relaxation 
rates $T_1^{-1}$ and $T_2^{-1}$, provided these are coupled. 
For the two-mode BEC considered here this is automatically the case
as given by Eqn.~(\ref{eqn-rel-times}). Thus we also expect a maximum
of the steady state value of the phase coherence (\ref{eqn-alpha-def})
for a finite value of $T_1^{-1}$.

Let us now determine the steady state value of the contrast (\ref{eqn-alpha-def}) 
which quantifies the phase coherence of the two wells, as a function of the 
system parameters and the relaxation rates. Obviously, the only steady 
state in the strict sense is given by $\vec s = \vec 0$ and $n=0$, 
corresponding to a completely empty trap. 
However, the system rapidly relaxes to a quasi-steady state where the 
internal dynamics is completely frozen out and all components of the 
Bloch vector and the particle number decay at the same rate:
\be
  \vec s(t) = \vec s_0 e^{-\kappa t}, \quad n(t) = n_0 e^{-\kappa t}.
  \label{eqn-static-ansatz}
\ee
Substituting this ansatz into the equations of motion (\ref{eqn-eom-bloch}),
the quasi-steady state is determined by the eigenvalue equation
\be
  \mathbf{M} 
  \begin{pmatrix}  s_{x0} \\ s_{y0} \\ s_{z0} \\ n_0  \end{pmatrix}
  = \kappa \begin{pmatrix}  s_{x0} \\ s_{y0} \\ s_{z0} \\ n_0  \end{pmatrix}
  \label{eqn-dic-eigen}
\ee
with the matrix
\be
  \mathbf{M} = \begin{pmatrix} 
   T_2^{-1} & 2\epsilon & 0 & 0 \\
   -2\epsilon & T_2^{-1} & -2J & 0 \\
   0 & 2J & T_1^{-1} & f_a T_1^{-1} \\
   0 & 0 & f_a T_1^{-1} & T_1^{-1}
 \end{pmatrix},
\ee
which is readily solved numerically.

\begin{figure}[tb]
\centering
\includegraphics[width=8cm, angle=0]{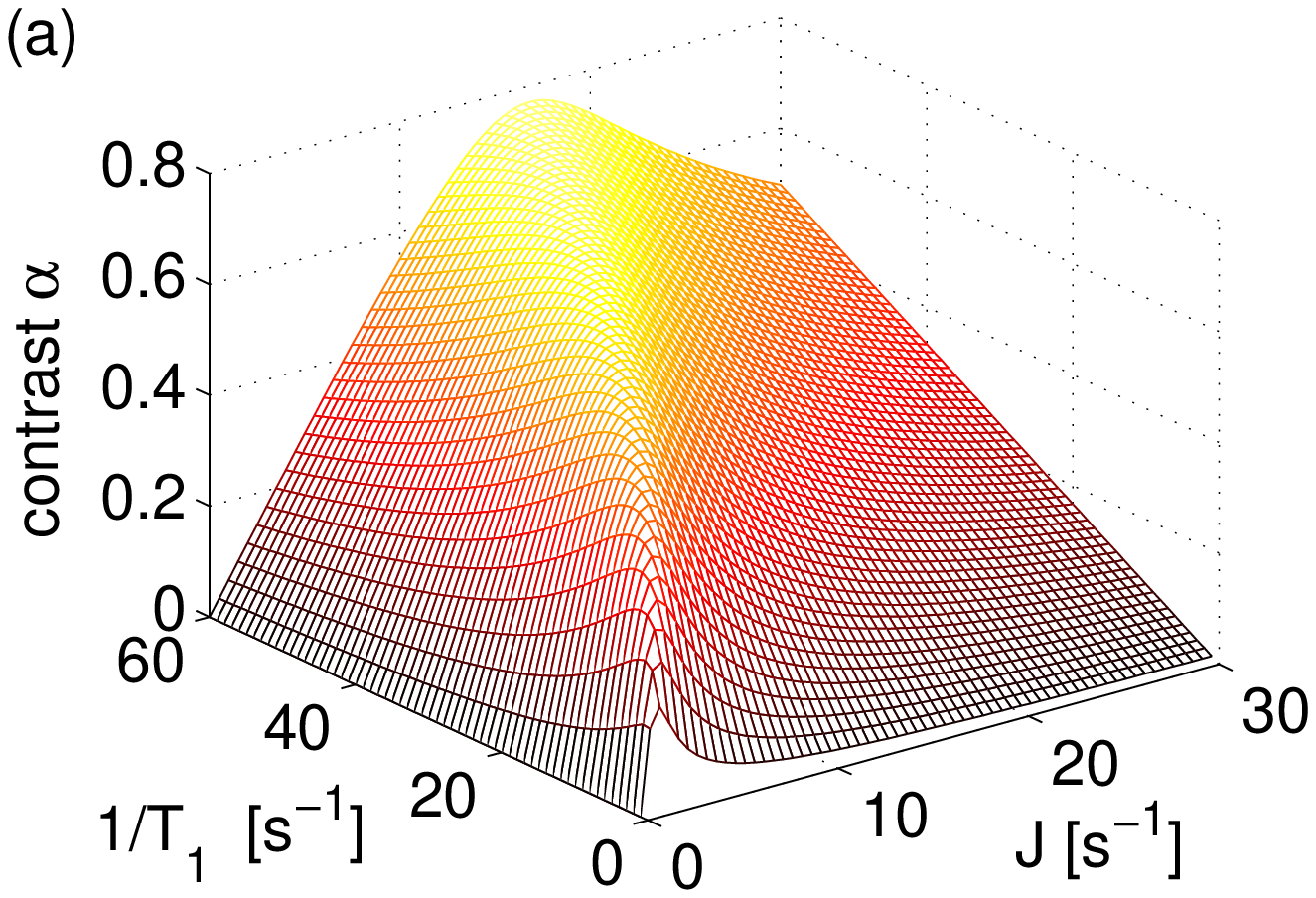}
\vspace{5mm}
\includegraphics[width=8cm, angle=0]{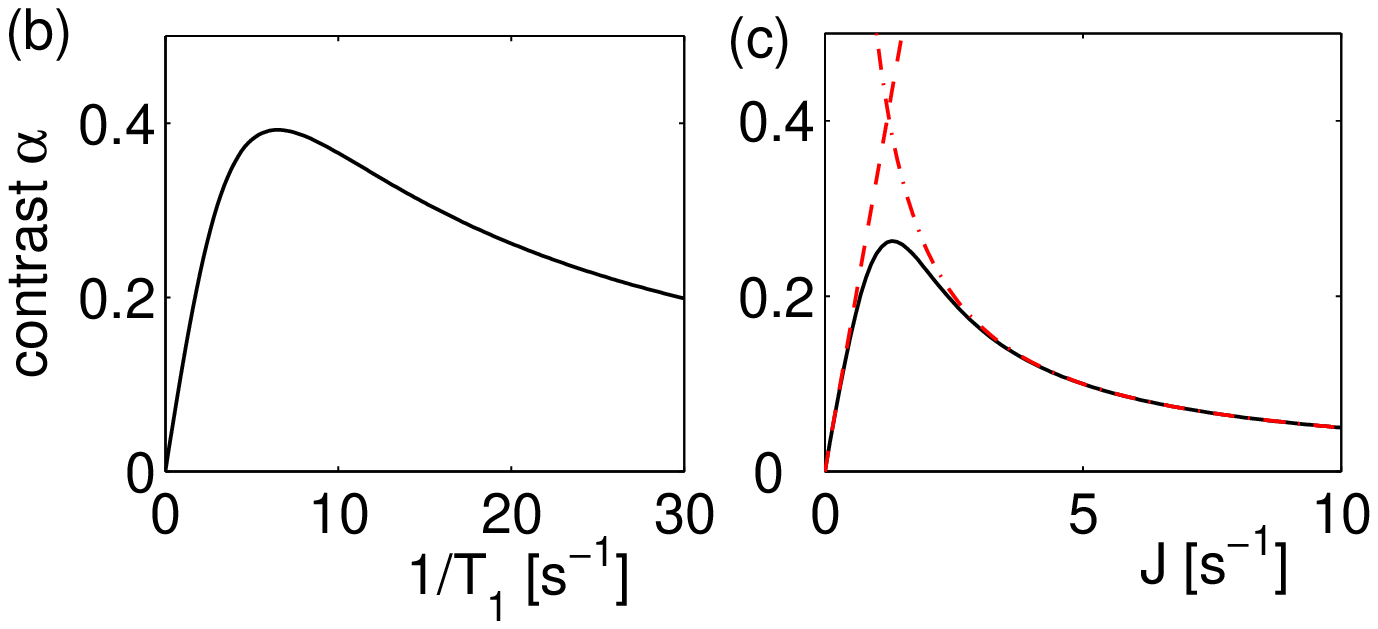}
\caption{\label{fig-contrast-3d}
(Color online) Contrast $\alpha$ of the quasi-steady state (\ref{eqn-static-ansatz}) 
as a function of the tunneling rate $J$ and the dissipation rate 
$1/T_1$ (a) for $\gamma_p = 5 \, {\rm s}^{-1}$ and $U = \epsilon = 0$ 
and (b) for a fixed value of the tunneling rate $J = 2 \, {\rm s}^{-1}$ 
and (c) a fixed value of the dissipation rate $1/T_1 = 2 \, {\rm s}^{-1}$.
The dash-dotted red lines represent the approximations (\ref{eqn-alpha-approx})
for small and large values of $J$, respectively.
}
\end{figure}

Fig.~\ref{fig-contrast-3d} shows the resulting values of the contrast $\alpha$
as a function of the dissipation rate $T_1^{-1}$ and the tunneling
rate $J$ for $U=\epsilon=0$ and $\gamma_p = 5 \, {\rm s}^{-1}$.
For a fixed value of one of the parameters, say $J$, one observes 
a typical SR-like maximum of the contrast for a finite value of 
the dissipation rate $1/T_1$ as shown in part (b) of the 
figure. In particular, the contrast is maximal if the time scales 
of the tunneling and the dissipation are matched according to
\be
  4J^2 \approx f_a^2 T_1^{-2} + f_a \gamma_p T_1^{-1}.
  \label{eqn-sr-condition}
\ee
Furthermore, the contrast $\alpha(J)$ shows a similar maximum for a 
finite value of the tunneling rate $J$ when the dissipation rate is
fixed as shown in Fig.~\ref{fig-contrast-3d} (c). Contrary to our
intuition this shows that an increase of the coupling of two modes 
can indeed \textit{reduce} their phase coherence.

In the special case $\epsilon=0$, illustrated in Fig.~\ref{fig-contrast-3d},
one can solve the eigenvalue equation (\ref{eqn-dic-eigen}) exactly.
In this case one has $s_x= 0$ and the contrast $\alpha$ is related to the 
eigenvalue $\kappa$ by 
\be
  \alpha = \frac{2J (T_1^{-1} - \kappa)}{f_a T_1^{-1} (T_2^{-1} - \kappa) }.
\ee
Evaluating the roots of the characteristic polynomial to determine $\kappa$
leads to an algebraic equation of third order which can be solved analytically.
The resulting expressions are quite lengthy, but the limits for small and
large values of the tunneling rate are readily obtained as
\bea
  & \alpha \approx \frac{2 J}{T_2^{-1} - (1- f_a) T_1^{-1}} \qquad & 
       \mbox{for} \, J \ll T_1^{-1} \nn \\
  & \alpha \approx \frac{f_a T_1^{-1}}{2 J}                 \qquad & 
         \mbox{for} \, J \gg T_1^{-1}.
  \label{eqn-alpha-approx}
\eea
These approximations are plotted as dashed red lines in 
Fig.~\ref{fig-contrast-3d} (c). Their intersection given by 
(\ref{eqn-sr-condition}) gives a very good approximation for the 
position of the SR-like maximum of the contrast $\alpha(J)$. 

\begin{figure}[tb]
\centering
\includegraphics[width=8cm, angle=0]{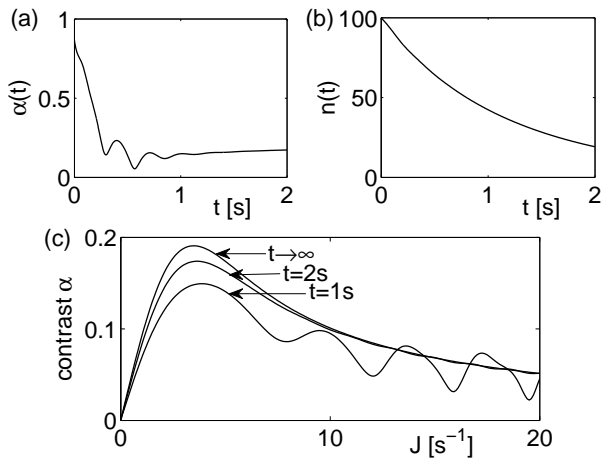}
\caption{\label{fig-relaxation}
(Color online)
Relaxation to the quasi-steady state 
for $\gamma_p = 5 \, {\rm s}^{-1}$, $T_1^{-1} = 1 \, {\rm s}^{-1}$, 
$\epsilon = 10 \, {\rm s}^{-1}$ and $U=0$. 
(a) Relaxation of the contrast $\alpha(t)$ for $J = 4 \, {\rm s}^{-1}$.
(b) Decay of the particle number $n(t)$ for $J = 4 \, {\rm s}^{-1}$.
(c) Development of the SR-maximum of the contrast $\alpha(J)$.
}
\end{figure}

An important experimental issue is the question whether the quasi-steady 
state is reached fast enough, such that the typical SR-like curve of the 
contrast as shown in Fig.~\ref{fig-contrast-3d}, can be observed while 
still enough atoms are left in the trap. To answer this question, we 
integrate the equations of motion (\ref{eqn-eom-bloch}) starting
from a pure BEC with $\vec s(0)/n(0) = (\sqrt{3}/2,0,1/2)$ 
and $n(0)=100$ particles.
Fig.~\ref{fig-relaxation} (a) shows the relaxation of the 
contrast for $J = 4 \, {\rm s}^{-1}$ and $T_1 = 1 \, {\rm s}$. 
The steady state value is nearly reached after $t = 1 \, {\rm s}$, 
when still $40 \%$ of the atoms are left in the trap.
Fig.~\ref{fig-relaxation} (b) shows the development of the
contrast $\alpha(J)$ in time. It is observed that the characteristic 
SR-like maximum is already well developed after 1 second, where
roughly half of the atoms are lost. Thus we conclude that the 
SR-like maximum of the contrast should be observable in 
ongoing experiments.
   
\begin{figure}[tb]
\centering
\includegraphics[width=7cm, angle=0]{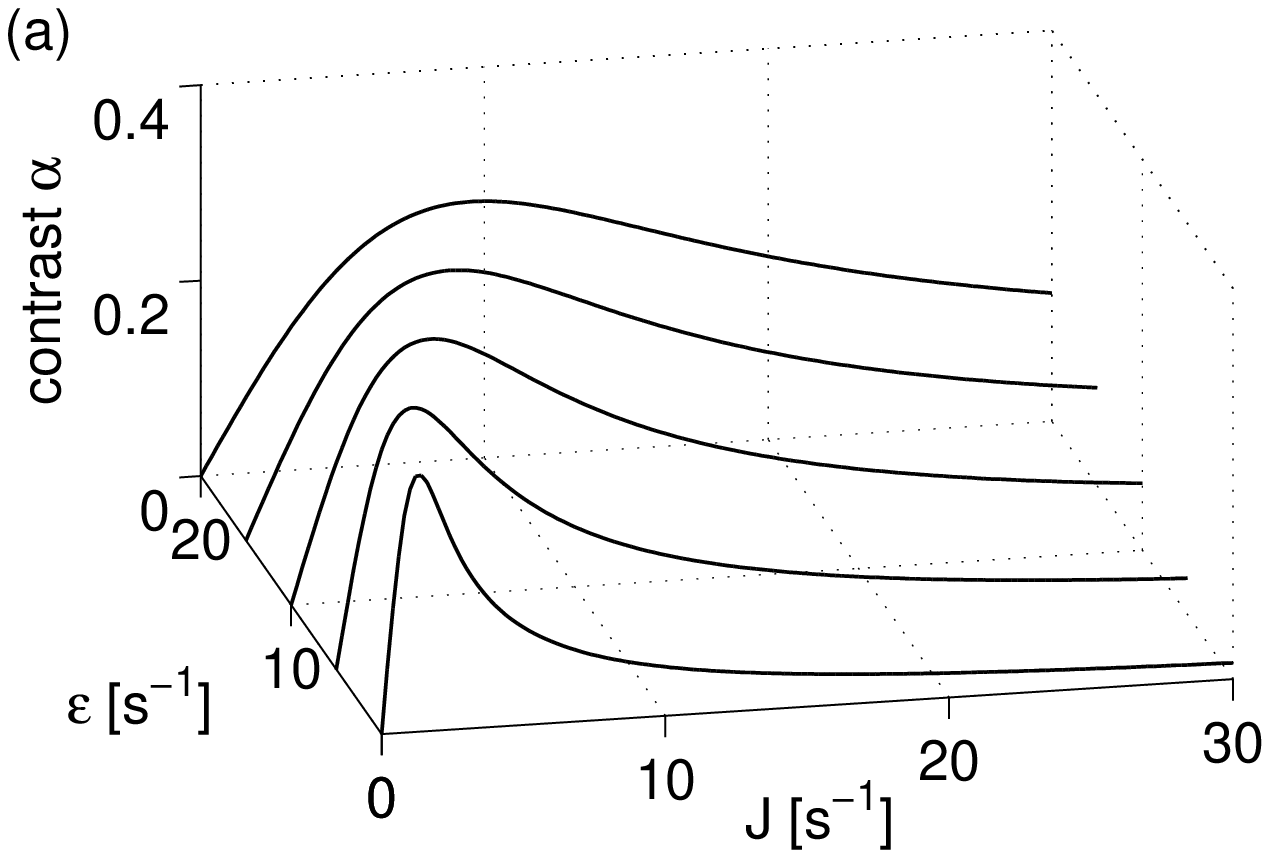}
\includegraphics[width=7cm, angle=0]{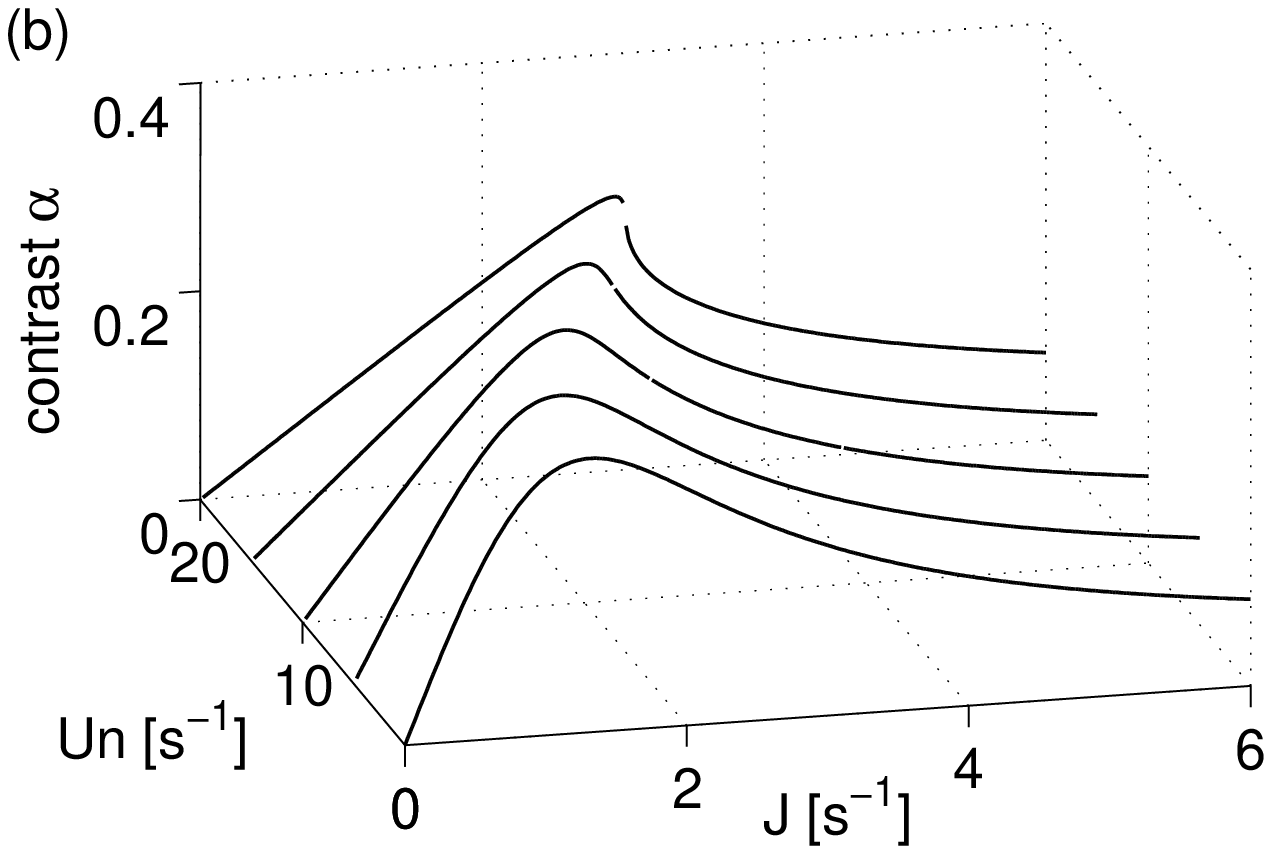}
\caption{\label{fig-contrast-jeg}
Steady state values of the contrast $\alpha$ as a function of 
the tunneling rate, for $U=0$ and different values of the energy 
bias $\epsilon$ (a), and as a function of the effective interaction 
strength $g=Un$ for $\epsilon=0$  (b). The remaining 
parameters are $\gamma_p = 5 \, {\rm s}^{-1}$ and 
$T_1^{-1} = 2 \, {\rm s}^{-1}$.}
\end{figure}

The stochastic resonance effect introduced above is robust and
generally not altered by changes of the system parameters or in the
presence of weak inter-particle interactions. For instance, a change 
of the bias $\epsilon$ of the on-site energies of the two wells preserves 
the general shape of $\alpha(1/T_1,J)$ shown in Fig.~\ref{fig-contrast-3d},
and especially the existence of a pronounced SR-like maximum. At most,
the function $\alpha(1/T_1,J)$ is stretched, shifting the position 
of the SR-like maximum. This shift is illustrated in Fig.~\ref{fig-contrast-jeg} (a)
where we have plotted the contrast as a function of $J$ for the 
dissipation rate $T_1^{-1} = 2 \, {\rm s}^{-1}$ and different values
of $\epsilon$. Thus, this effect provides a useful tool to shift the maximum 
to values of $J$, which are easier accessible in ongoing experiments.

Similarly, the position of the maximum of the coherence $\alpha(J)$
is shifted in the presence of weak inter-particle interactions. 
An interacting BEC will usually not show a simple 
exponential decay of the form (\ref{eqn-static-ansatz}) because
the instantaneous decay rate depends on the effective interaction 
strength $Un(t)$, which also decreases \cite{Wimb05,Schl06,Schl07}. However, the 
discussion of quasi-steady states and instantaneous decay rates is
still useful if the decay is weak. In this case the system can follow 
the quasi-steady states adiabatically and the decay of the population is
given by 
\bea
  \frac{d n(t)}{dt} &=& - \kappa(n(t)) n(t) \qquad \mbox{and} \nn \\ 
  \frac{d \vec s(t)}{dt} &=& - \kappa(n(t)) \vec s(t)
  \label{eqn-nonlin-adia}
\eea
in good approximation. Substituting this ansatz into the equations
of motion (\ref{eqn-eom-bloch}) yields four
coupled nonlinear algebraic equations, which can be disentangled
with a little algebra. For a given number of particles $n$, the 
instantaneous decay rate $\kappa$ is obtained by solving the fourth order
algebraic equation
\begin{align}
  &\left[(\kappa - T_2^{-1})^2 + (Un)^2 (\kappa - T_1^{-1})^2 \right]
  \left[(\kappa - T_1^{-1})^2  - f_a^2 T_1^{-2} \right] \nn \\
  & \qquad + 4J^2 f_a^2 T_1^{-2} (\kappa - T_1^{-1}) (\kappa - T_2^{-1}) = 0.
  \label{eqn-static-nonlin}
\end{align}
The Bloch vector for the corresponding quasi-steady state is then given by
\bea
  s_{x0} &=& \frac{\kappa - T_1^{-1}}{\kappa - T_2^{-1}}
     \frac{(\kappa - T_1^{-1})^2  - f_a^2 T_1^{-2}}{2J f_a^2 T_1^{-2}} Un^2 \nn \\
  s_{y0} &=& \frac{(\kappa - T_1^{-1})^2  - f_a^2 T_1^{-2}}{2 J f_a T_1^{-1}} n \nn \\
  s_{z0} &=& \frac{\kappa - T_1^{-1}}{f_a T_1^{-1}} n. \nn \\
\eea
The fourth order equation (\ref{eqn-static-nonlin}) yields four solutions
for the decay rate $\kappa$. Discarding unphysical values, one finds either 
one or three quasi-steady states. This appearance of novel nonlinear stationary 
states has been discussed in detail in the context of nonlinear Landau-Zener 
tunneling \cite{Wu00,Wu06,06LZnonlin1,06LZnonlin2} and nonlinear
transport \cite{Paul05,06nl_transport}.

The resulting contrast $\alpha(J)$ in a quasi-steady state is shown in 
Fig.~\ref{fig-contrast-jeg} (b) for different values of the effective 
interaction constant $g=Un$. One observes that the position of the 
SR-like maximum of the contrast is shifted to larger values of the 
tunneling rate, while the height remains unchanged.
Furthermore the shape of the stochastic resonance curve $\alpha(J)$
is altered, becoming flatter for $J < J_{\rm max}$ and steeper for
$J > J_{\rm max}$. 
For even larger values of the interaction constant $Un$ one finds a 
bifurcation into three distinct quasi-steady states as introduced above.
This case will be discussed in detail in Sec.~\ref{sec-dic-strong}.

\begin{figure}[tb]
\centering
\includegraphics[width=8cm, angle=0]{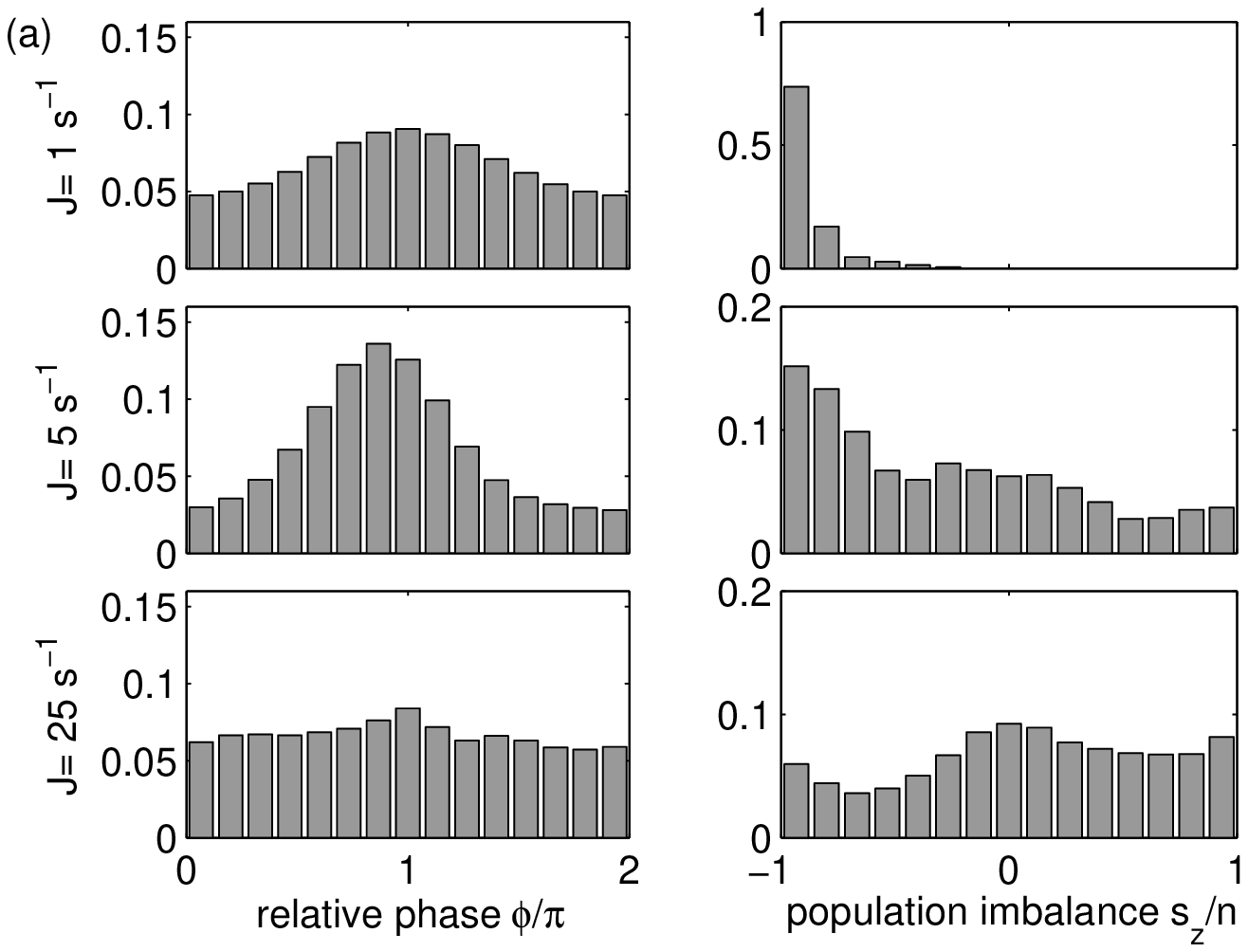}
\vspace{5mm}
\includegraphics[width=8cm, angle=0]{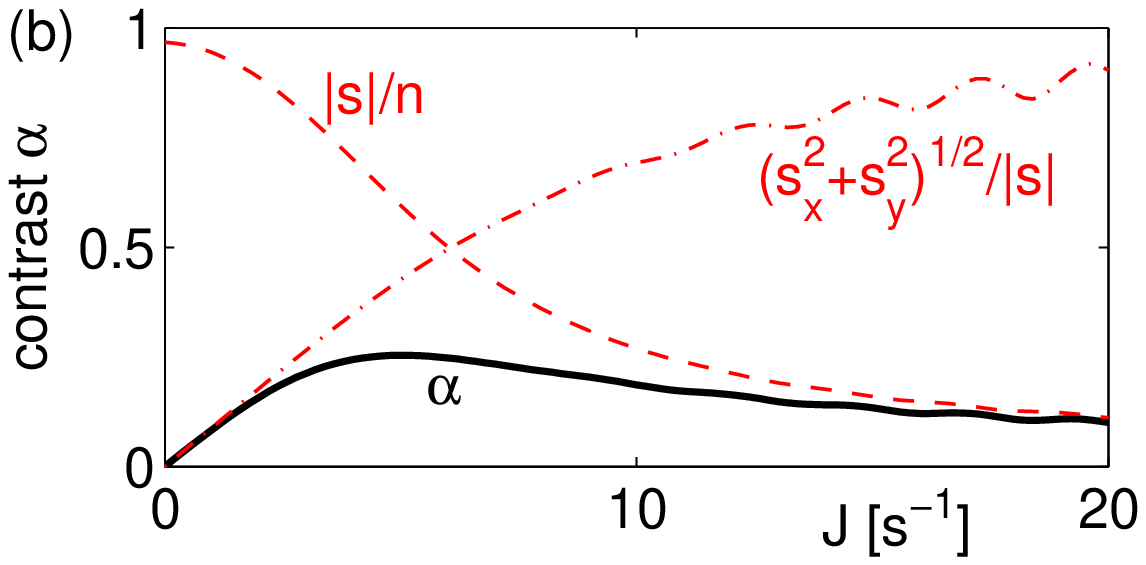}
\caption{\label{fig-qjump-lin}
(Color online)
(a) Histogram of the probabilities to measure the relative phase $\phi$
and the relative population imbalance $s_z$ in a single experimental 
run after $t = 1.5 \, {\rm s}$ obtained from a MCWF simulation of the 
many-body dynamics. The initial state was chosen as a pure BEC
(i.e.~a product state) with $s_z = n/2$ and $n(0) = 100$ particles
and the remaining parameters are $\gamma_p = 5 \, {\rm s}^{-1}$,
$T_1 = 0.5 \, {\rm s}$, $\epsilon = 10 \, {\rm s}^{-1}$, $U=0.1 \, {\rm s}^{-1}$.
(b) Average contrast $\alpha = \sqrt{s_x^2+s_y^2}/n$  (solid black line) 
after $t = 1.5 \, {\rm s}$ compared to $\sqrt{s_x^2+s_y^2}/|\vec s|$ 
and $|\vec s|/n$ (dashed red lines).
}
\end{figure} 

The reasons for the occurrence of an SR-like maximum of the contrast 
in terms of the underlying many-particle dynamics are illustrated in 
Fig.~\ref{fig-qjump-lin}. 
To obtain these results we have simulated the dynamics generated
by the Master equation (\ref{eqn-master2}) using the Monte Carlo 
wave function (MCWF) method  \cite{Dali92,Molm93,Carm93} averaging over 
100 quantum trajectories. 
For a given particle number $n$, the probabilities $P$ to obtain
the population imbalance $s_z$ and the relative phase $\phi$
in a projective measurement are thereby given by
\bea
  P(s_z)  &=& \tr( \ket{s_z}\bra{s_z}   \hat \rho) \qquad \mbox{and} \nn \\
  P(\phi) &=& \tr( \ket{\phi}\bra{\phi} \hat \rho),
  \label{eqn-szphi-prob}
\eea
where the $\hat L_z$ eigenstates 
\bea
  && \ket{s_z} = \ket{n/2-s_z,n/2+s_z}  \quad \mbox{with} \nn \\
  && s_z = -n/2,-n/2+1,\ldots, n/2
\eea
and the phase eigenstates
\bea
   && \ket{\phi} := \frac{1}{\sqrt{n+1}} \sum_{s_z=-n/2}^{+n/2}  e^{i \phi s_z} \ket{s_z}
       \quad \mbox{with}  \nn \\
   && \phi = 0, 2\pi \frac{1}{n+1}, \ldots, 2\pi \frac{n}{n+1}
\eea
each form a complete basis.

Part (a) of Fig.~\ref{fig-qjump-lin} shows a histogram of the probabilities 
to observe the relative population imbalance $s_z/n$ and the relative 
phase $\phi$ in a single experimental run for three different values 
of the tunneling rate $J$ after the system has relaxed to the 
quasi-steady state. With increasing $J$, the atoms are distributed 
more equally between the two wells so that the single shot contrast 
increases. Within the mean-field description this is reflected
by an increase of $\sqrt{s_x^2+s_y^2}/|\vec s|$ at the expense of 
$|s_z|/|\vec s|$ (cf. part (b) of the figure). However, this effect also makes 
the system more vulnerable to phase noise so that the relative phase 
of the two modes becomes more and more random and $|\vec s|/n$ decreases. 
The average contrast (\ref{eqn-alpha-def}) then assumes a maximum for 
intermediate values of $J$ as shown in part (b) of the figure.

\section{Stochastic resonance of a driven BEC}
\label{sec-sr-weak}

So far we have demonstrated a stochastic resonance of the contrast 
for a BEC in a static double-well trap with biased particle losses.
In the following we will show that the system's response to a weak 
external driving also assumes a maximum for a finite dissipation 
rate -- an effect which is conceptually closer to the common 
interpretation of stochastic resonance. 
From a mathematical viewpoint, however, one can rather relate the 
{\it undriven} case discussed above to the stochastic resonance effect 
in nuclear magnetic resonance \cite{Viol00}.
In fact, the Bloch equations for the magnetization have constant
coefficients in the rotating wave approximation, and should thus be 
compared to the undriven equations of motion (\ref{eqn-eom-bloch}).

\begin{figure}[tb]
\centering
\includegraphics[width=7cm, angle=0]{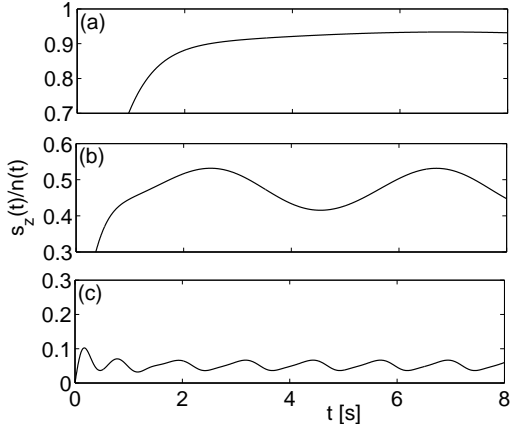}
\caption{\label{fig-response-ex}
Dynamics of the relative population imbalance $s_z(t)/n(t)$ in a weakly 
driven double well trap for three different values of the tunneling rate: 
$J_0 = 0.5 \, {\rm s}^{-1}$ (a), $J_0 = 1.5 \, {\rm s}^{-1}$ (b) and 
$J_0 = 5 \, {\rm s}^{-1}$ (c). The amplitude of the forced oscillations
is maximal for intermediate values of $J_0$ as shown in part (b).
The remaining parameters are $T_1^{-1} = 2 \, {\rm s}^{-1}$, $U=0$, 
$\epsilon=0$, $\gamma_p = 5 \, {\rm s}^{-1}$ and $J_1/J_0 = 10\%$.
Please note the different scalings.
}
\end{figure}

Let us consider the response of the system to a
weak sinusoidal driving of the tunneling rate
\be
  J(t) = J_0 + J_1 \cos(\omega t)
\ee  
at the resonance frequency $\omega = \sqrt{J_0^2 + \epsilon^2}$, while 
the amplitude of the driving is small and fixed as $J_1/J_0 = 10\%$. 
A variation of $J$ can be realized in a quite simple way in an optical 
setup \cite{Albi05,Gati06a,Gati06b}, 
where the tunneling barrier between the  two wells is given by an optical 
lattice formed by two counter-propagating laser beams. A variation of the 
intensity of the laser beams then directly results in a variation of the tunneling 
rate $J$. 
Fig.~\ref{fig-response-ex} shows the resulting dynamics for 
$T_1 = 0.5 \, {\rm s}$ and three different values of $J_0$ and $U=0$.
After a short transient period, the relative population 
imbalance $s_z(t)/n(t)$ oscillates approximately sinusoidally.
One clearly observes that the response, i.e. the amplitude of the 
forced oscillations, assumes a maximum for intermediate values of 
$J_0$ matching the external time scale of the dissipation given by 
$T_1^{-1}$. 

For a detailed quantitative analysis of this stochastic resonance effect, we evaluate 
the amplitude of the oscillation based on a linear response argument for $U=0$. In the following, 
we will use a complex notation for notational convenience, while only the real part is 
physically significant.
The equations of motion (\ref{eqn-eom-bloch}) are then rewritten in matrix form as
\be
  \frac{d}{dt} \begin{pmatrix} \vec s \\ n \end{pmatrix} =
  \left( \mathbf{M}_0 + \mathbf{M}_1 e^{i\omega t} \right) \begin{pmatrix} \vec s \\ n \end{pmatrix}. 
\ee
The matrices $M_0$ and $M_1$ are defined by 
\be
  \mathbf{M_0} = \begin{pmatrix} 
   T_2^{-1} & 2\epsilon_0 & 0 & 0 \\
   -2\epsilon_0 & T_2^{-1} & -2J_0 & 0 \\
   0 & 2J_0 & T_1^{-1} & f_a T_1^{-1} \\
   0 & 0 & f_a T_1^{-1} & T_1^{-1}
 \end{pmatrix}
\ee
and 
\be
  \mathbf{M_1} = \begin{pmatrix} 
   0 & 0 & 0 & 0 \\
   0 & 0 & -2J_1 & 0 \\
   0 & 2J_1 & 0 & 0 \\
   0 & 0 & 0 & 0
 \end{pmatrix}.
\ee
As before we consider the dynamics after all transient oscillations have died
out, assuming that $\vec s(t)$ as well as $n(t)$ decay exponentially at the same
rate. However, we now also have an oscillating contribution so that we make
the ansatz
\bea
  \vec s(t) &=& (\vec s_0 + \vec s_1 e^{i \omega t}) e^{-\kappa t}\nn \\
  n(t) &=& (n_0 + n_1 e^{i \omega t}) e^{-\kappa t}.
\eea  
The amplitude of the oscillations, i.e. the system response, is thus directly
given by $\vec s_1/n_0$. Substituting this ansatz in the equations of motion
and dividing by $e^{-\kappa t}$ yields
\bea
   && -\kappa \begin{pmatrix} \vec s_0 \\ n_0 \end{pmatrix}
   + (i\omega - \kappa) \begin{pmatrix} \vec s_1 \\ n_1 \end{pmatrix} e^{i\omega t} 
     \label{eqn-motion-driven1} \\
   && \quad = \left[ \mathbf{M}_0 + \mathbf{M}_1 e^{i\omega t} \right]
       \times \left[ \begin{pmatrix} \vec s_0 \\ n_0 \end{pmatrix} +
       \begin{pmatrix} \vec s_0 \\ n_0 \end{pmatrix} e^{i \omega t} \right]. \nn
\eea

\begin{figure}[tb]
\centering
\includegraphics[width=7cm, angle=0]{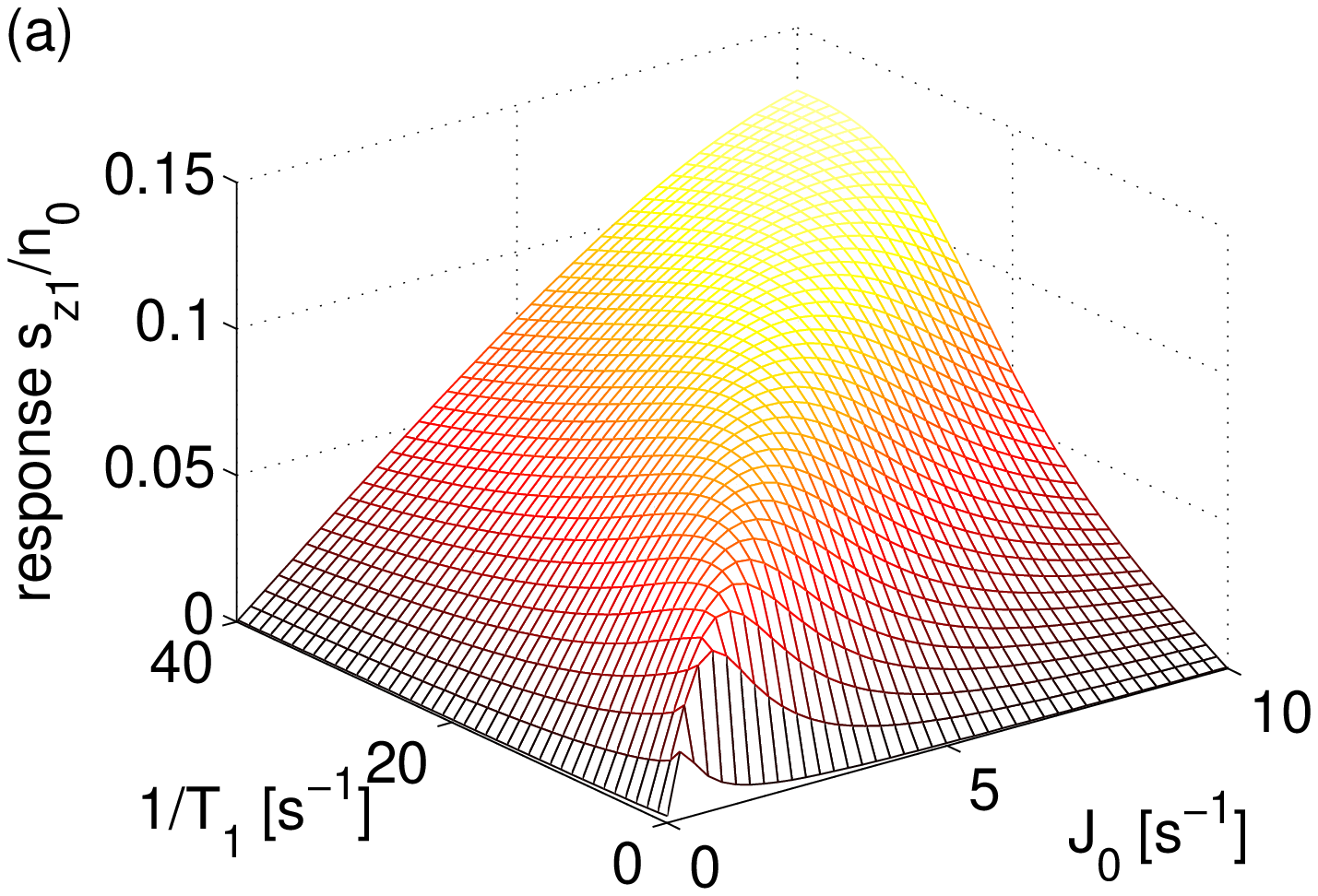}
\includegraphics[width=7cm, angle=0]{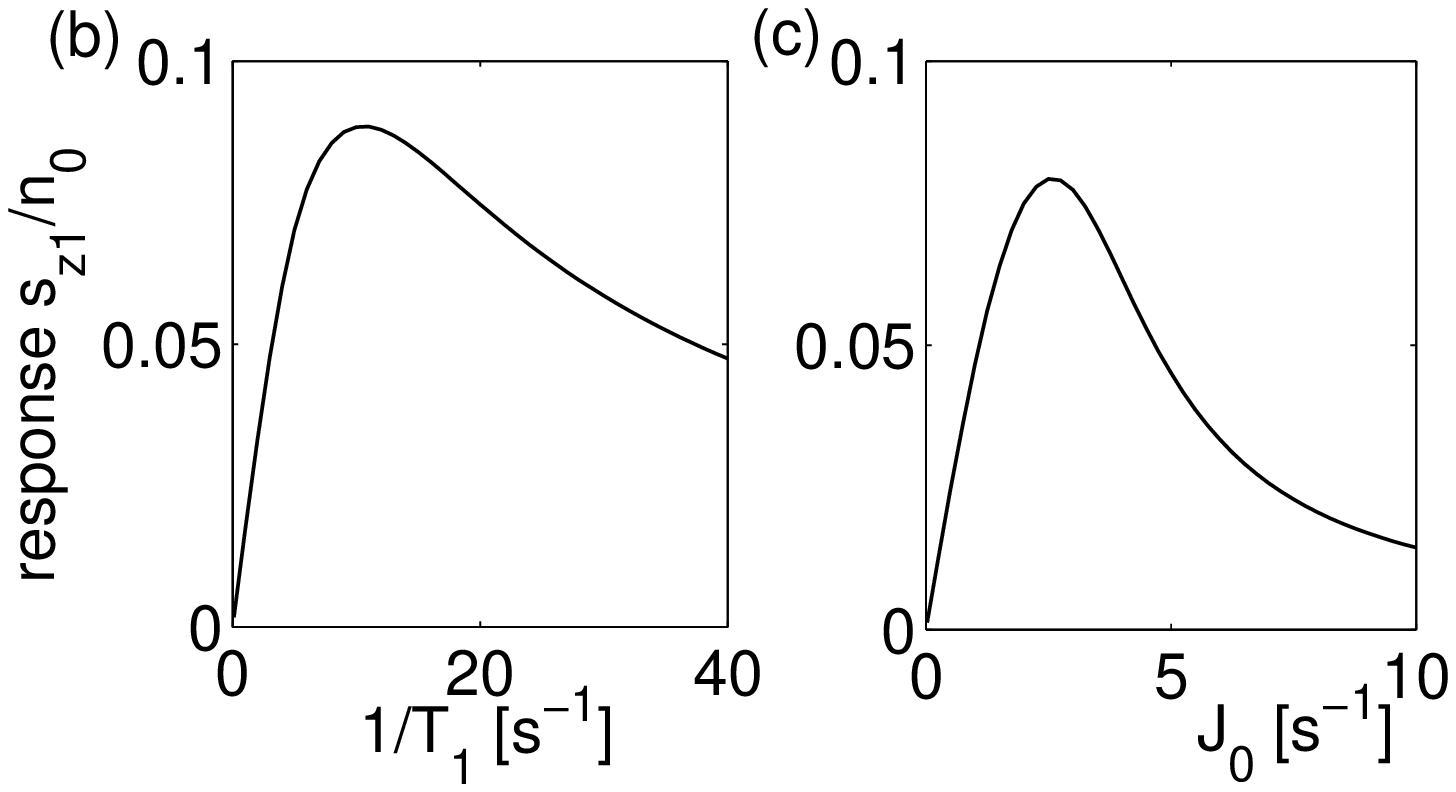}
\caption{\label{fig-response-dj}
(Color online)
(a) Response (amplitude of the oscillations of $s_z(t)/n(t)$) of
a weakly driven double well trap vs. $T_1^{-1}$ and
$J_0$ calculated within linear response theory. (b) For a fixed
value of the tunneling rate $J_0 = 2.5 \, {\rm s}^{-1}$.
(c) For a fixed value of the dissipation rate 
$T_1^{-1} = 2 \, {\rm s}^{-1}$. 
The remaining parameters are $U=0$, $\epsilon=0$, 
$\gamma_p = 5 \, {\rm s}^{-1}$ and $J_1/J_0 = 10\%$.
}
\end{figure}   

Neglecting the higher order terms $\sim e^{2 i \omega t}$ in a linear response 
approximation and dividing Eqn.~(\ref{eqn-motion-driven1}) in the time-dependent 
and the time-independent parts yields the equations
\be
  \left[ -\mathbf{M}_0 + (i\omega-\kappa) \eye \right] \begin{pmatrix} \vec s_1 \\ n_1 \end{pmatrix}  
  = \mathbf{M}_1 \begin{pmatrix} \vec s_0 \\ n_0 \end{pmatrix}
   \label{eqn-driven-s1n}
\ee
and (\ref{eqn-dic-eigen}), which determine $\vec s_1$ and $n_1$. The resulting 
values of the system response  are shown in Fig.~\ref{fig-response-dj}.
One observes the characteristic signatures of a stochastic resonance:
If one of the two parameters $J_0$ and $T_1$ is fixed, the response
assumes a maximum for a finite value of the remaining parameter as
shown in part (b) and (c) of the figure. Part (a) shows that this maximum
is assumed if the external ($T_1^{-1}$) and the internal ($J_0$) timescale are 
matched similar to the undriven case illustrated in Fig.~\ref{fig-contrast-3d}.
Let us stress that this scenario is again not fundamentally altered in the 
case of weak interactions as numerically tested but not shown here.

\begin{figure}[tb]
\centering
\includegraphics[width=7cm, angle=0]{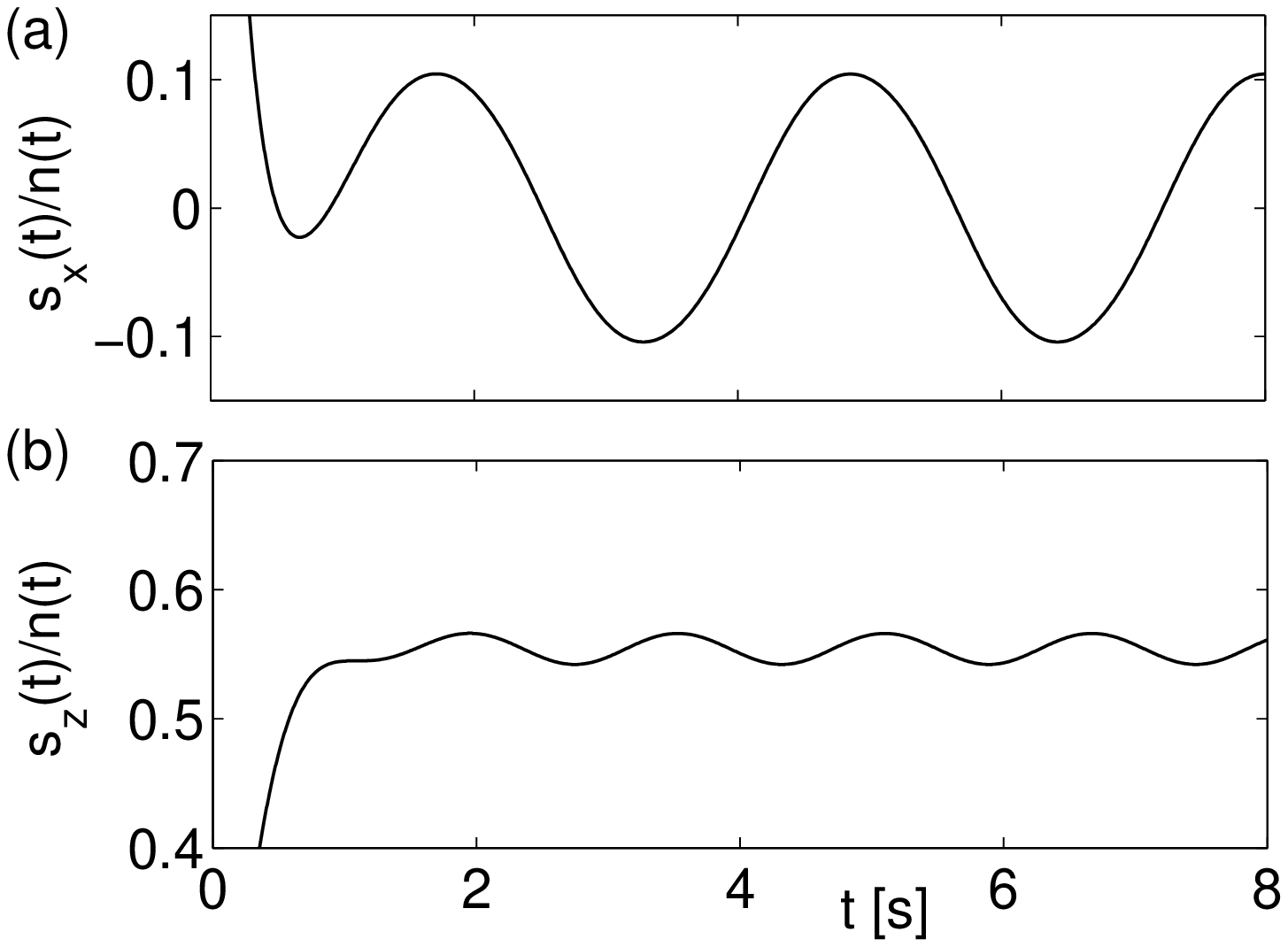}
\includegraphics[width=7cm, angle=0]{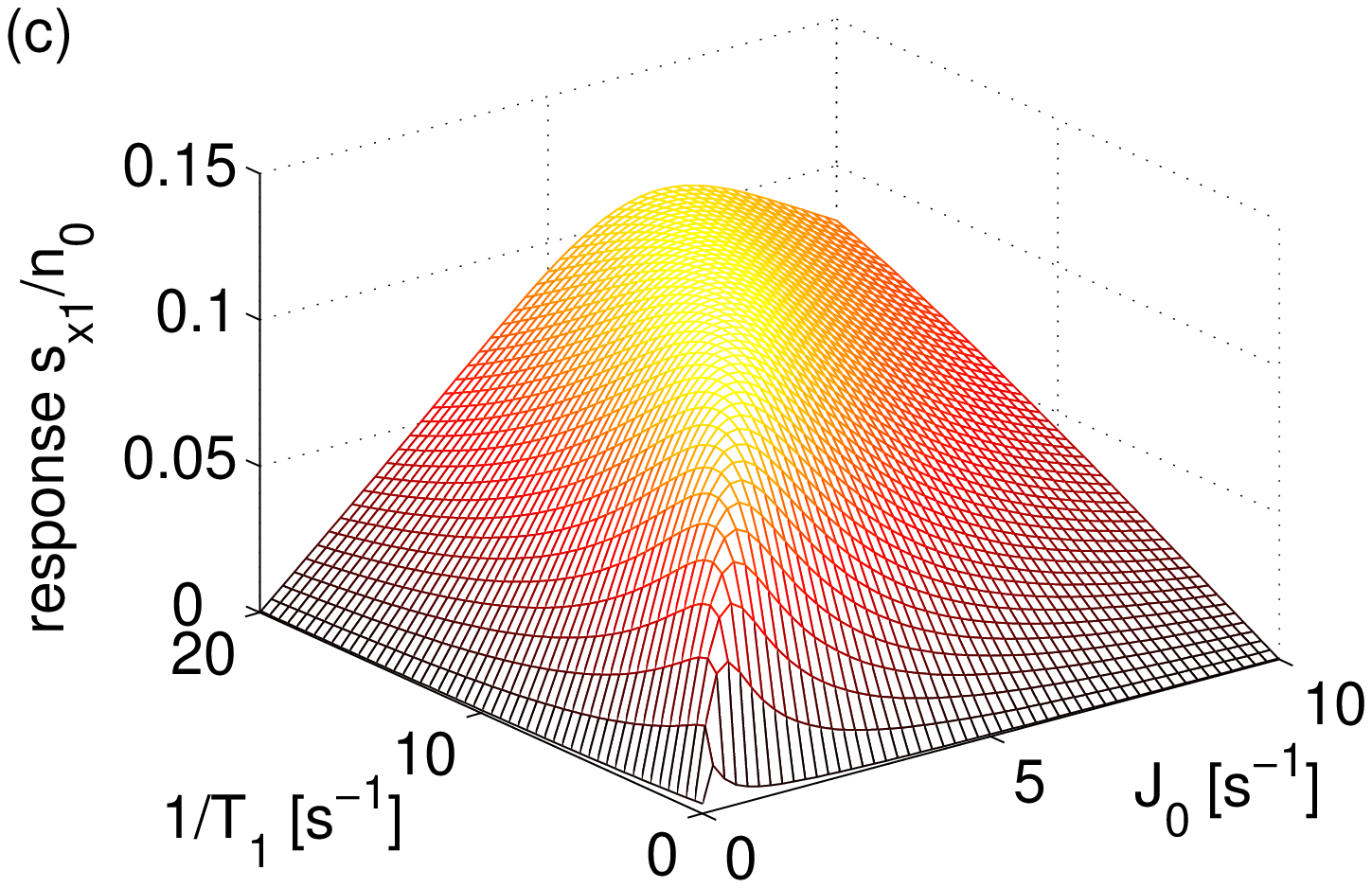}
\caption{\label{fig-driven-e1}
(Color online)
Dynamics of the coherence $s_x(t)/n(t)$ (a) and the relative 
population imbalance $s_z(t)/n(t)$ (b) for a double well trap
with a driven energy bias $\epsilon$ for $J_0 = 2 \, {\rm s}^{-1}$ 
and $T_1^{-1} = 4 \, {\rm s}^{-1}$.
(c) Response (amplitude of the oscillations of $s_x(t)/n(t)$) vs. 
$T_1^{-1}$ and $J_0$ calculated within linear response theory. 
The remaining parameters are $U=0$, $\epsilon_1=1 \, {\rm s}^{-1}$, 
$\gamma_p = 5 \, {\rm s}^{-1}$.
}
\end{figure}   

A different situation arises if the energy bias is driven
instead of the tunneling rate $J$ such that
\be
  \epsilon(t) = \epsilon_1 \cos( \omega t).
\ee
As above we can evaluate the amplitude of the forced oscillations
within the linear response theory, however with
\be
  \mathbf{M_1} = \begin{pmatrix} 
   0 & -2\epsilon_1 & 0 & 0 \\
   2\epsilon_1 & 0 & 0 & 0 \\
   0 & 0 & 0 & 0 \\
   0 & 0 & 0 & 0
 \end{pmatrix}.
\ee
Solving the equations (\ref{eqn-driven-s1n}) and (\ref{eqn-dic-eigen})
then yields $s_{1y} = s_{1z} = 0$. Remarkably, a driving of the energy 
bias does not affect the population imbalance in leading order. 
Only the first component of the Bloch vector $s_x$, and thus also 
the contrast $\alpha$ is strongly affected.

This is illustrated in Fig.~\ref{fig-driven-e1} (a) and (b) where the 
relative population imbalance $s_z(t)/n(t)$ and the first component 
of the Bloch vector $s_x(t)/n(t)$ are plotted for $J_0 = 2 \, {\rm s}^{-1}$, 
$T_1^{-1} = 4 \, {\rm s}^{-1}$ and $\epsilon_1=1 \, {\rm s}^{-1}$.
The coherence oscillates strongly at the fundamental frequency $\omega$,
while the population imbalance oscillates only with a tiny amplitude at 
the second harmonic frequency $2\omega$.
The oscillation amplitude of the coherence then again shows the familiar
SR-like dependence on the parameters $J_0$ and $T_1$ as illustrated
in Fig.~\ref{fig-driven-e1} (c).

\begin{figure}[tb]
\centering
\includegraphics[width=8cm, angle=0]{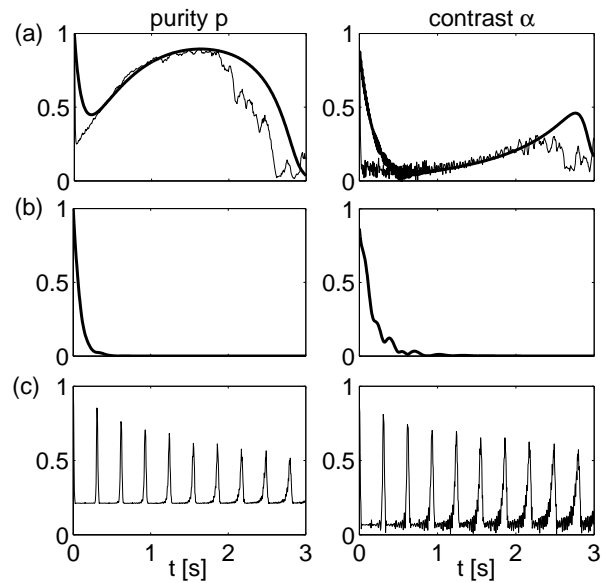}
\caption{\label{fig-strong-pa1}
(a) Time evolution of the purity $p$ and the contrast $\alpha$ for
$J=U=10 \, {\rm s}^{-1}$, $\epsilon= 0$, $T_1 = 0.5 \, {\rm s}$.
(b) Time evolution without interactions ($U=0$) and (c) without
dissipation ($1/T_1 = 1/T_2 = 0$) for comparison. The occasional 
revivals are artifacts of the small particle number.
The initial state is a pure BEC  with $s_z = n/2$ and 
$n(0) = 100$ particles. 
The results of a MCWF simulation averaged over 100 runs are 
plotted as a thin solid line in (a) and (c), while the mean-field 
results are plotted as a thick line in (a) and (b).
Note that the mean-field approximation is exact in case (b),
whereas it breaks down in case (c) and is thus not shown
(cf. \cite{Vard01,08phase2}).}
\end{figure}

\section{Dissipation induced coherence in a strongly-interacting BEC}
\label{sec-dic-strong}

Let us finally discuss the case of strong interactions, which 
is experimentally most relevant and theoretically most 
profound. An example for the dynamics of a strongly-interacting 
BEC is shown in Fig.~\ref{fig-strong-pa1} (a) for an initially 
pure BEC with $s_z = n/2$, calculated both with the MCWF method 
and within the mean-field approximation (\ref{eqn-eom-bloch}). 
One observes that the purity $p$ and the contrast $\alpha$ first 
drop rapidly due to the phase noise and, more importantly, due to
the interactions. This is an effect well-known from the non-dissipative 
system and can be attributed to a dynamical instability which also leads 
to the breakdown of the mean-field approximation 
\cite{Vard01,Angl01,08phase2,Cris04}. 
However, a surprising effect is found at intermediate times:
The purity $p$ is restored almost completely and the contrast 
$\alpha$ is slightly increasing. 

\begin{figure}[tb]
\centering
\includegraphics[width=4cm, angle=0]{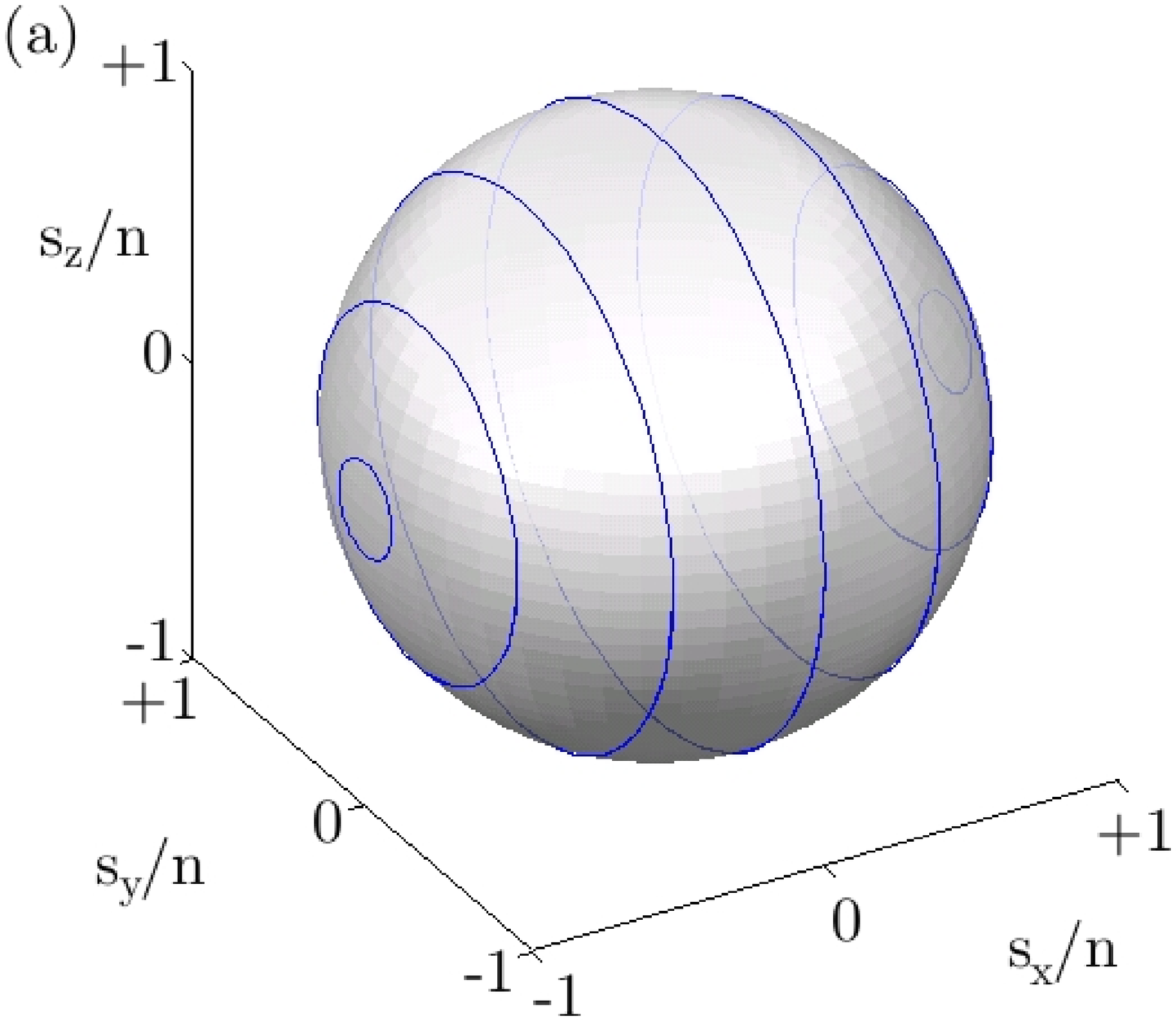}
\includegraphics[width=4cm, angle=0]{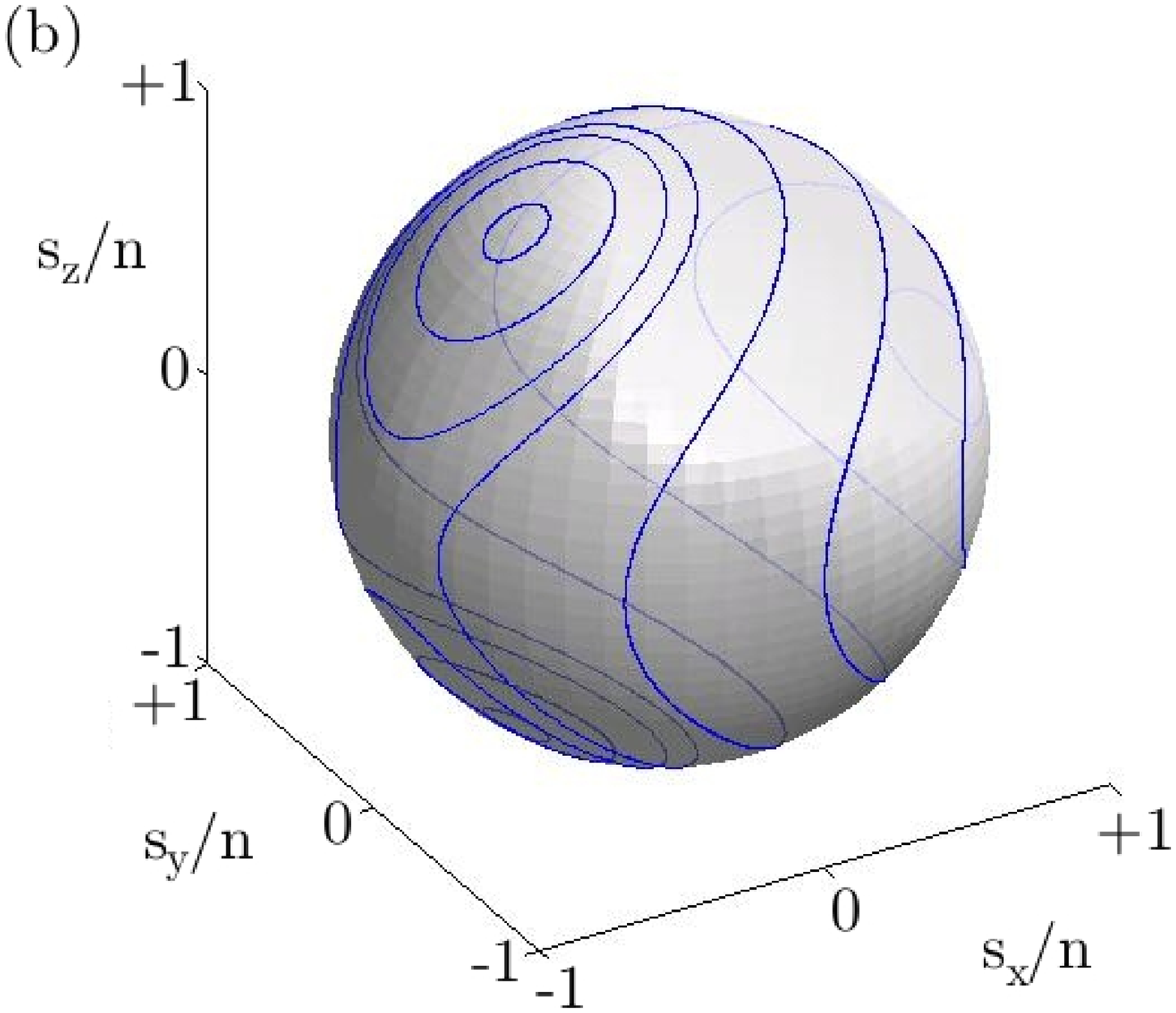}
\includegraphics[width=4cm, angle=0]{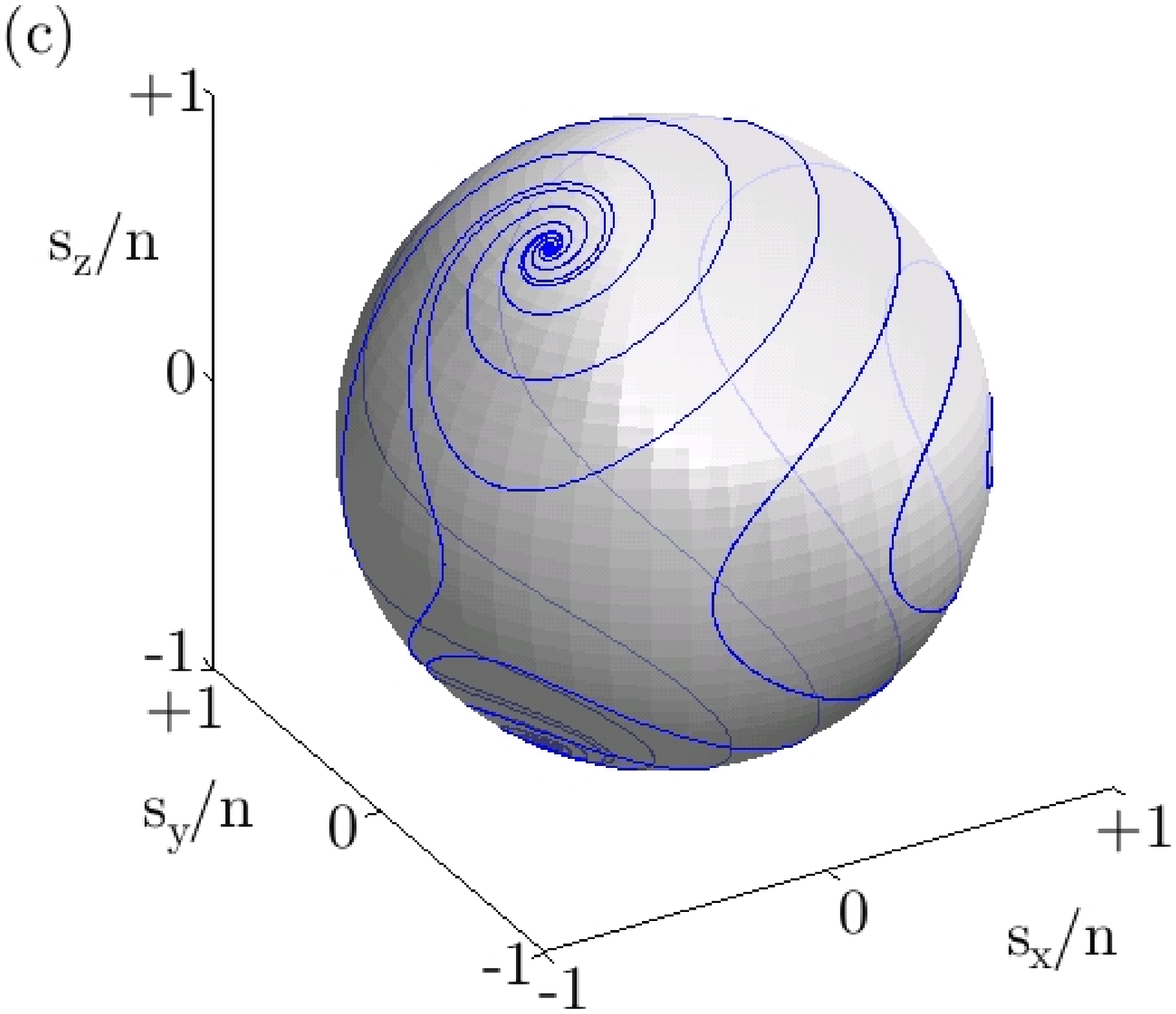}
\caption{\label{fig-spheres}
(Color online)
Mean-field dynamics without interactions and dissipation (a),
with interactions $Un = 40 \,{\rm s}^{-1}$ (b) and with 
interactions and dissipation $\gamma_a=10 \, {\rm s}^{-1}$ (c).
The remaining parameters are $J=10 \, {\rm s}^{-1}$ and $\epsilon =0$.
To increase the visibility we have plotted the rescaled Bloch vector
$\vec s/n$ and we have artificially fixed the particle number so
that $n = const$.
}
\end{figure}

Most interestingly, the observed values of the purity and the coherence
are much larger than in the cases where one of the two effects --
interactions and dissipation -- is missing. The time evolution for 
these two cases are also shown in Fig.~\ref{fig-strong-pa1}.
In the case of no interactions both purity and coherence rapidly 
drop to values of almost zero and do not revive. This case has
been discussed in detail in Sec.~\ref{sec-dic-weak}.
In the interacting case without dissipation one observes regular 
revivals, which are artifacts of the small particle number in 
the simulation and become less pronounced with increasing particle 
number. Apart from these occasional revivals, however, the 
purity and the coherence relax to values which are much smaller
than in the interacting {\it and} dissipative case.

The surprising re-purification of a strongly-interacting BEC by
particle dissipation can be understood within a semiclassical
phase space picture. In order to visualize the effects of particle 
loss, we have plotted the 'classical' phase space structure generated 
by the Bloch equations (\ref{eqn-eom-bloch}) for $\gamma_p = 0$
in Fig.~\ref{fig-spheres} without interactions and dissipation (a), 
with interactions (b) and with both (c). 
For illustrative purposes, we have plotted the rescaled Bloch vector
$\vec s/n$ and have artificially fixed the particle number so
that $n = const$. Since we are interested only in the short-time 
dynamics of the Bloch vector and not in the decay of the particle 
number on longer time scales, this is an appropriate treatment.
Moreover, in the quantum jump picture this approximation corresponds 
to the periods of constant particle number between two loss processes  
\cite{Dali92,Carm93,08mfdecay}.

\begin{figure}[tb]
\centering
\includegraphics[width=8cm, angle=0]{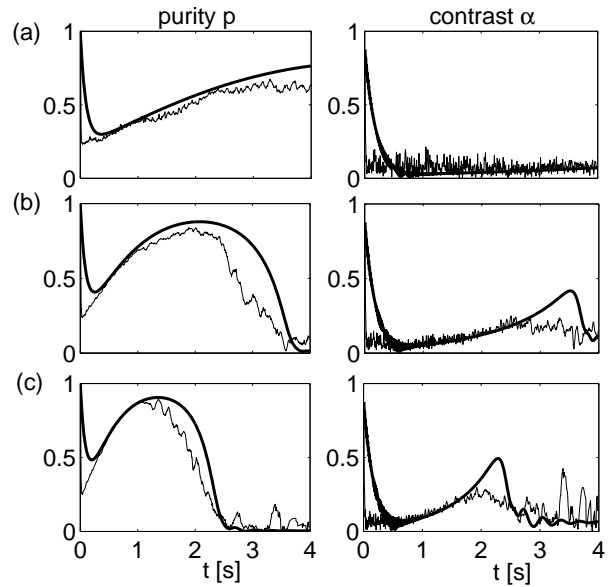}
\caption{\label{fig-strong-pa2}
Time evolution of the purity $p$ and the contrast $\alpha$ for
$J=U=10 \, {\rm s}^{-1}$, $\epsilon= 0$ and 
$1/T_1 = 0.5 \, {\rm s}^{-1}$ (a),
$1/T_1 = 1.5 \, {\rm s}^{-1}$ (b) and
$1/T_1 = 2.5 \, {\rm s}^{-1}$ (c).
The initial state is a pure BEC  with $s_z = n/2$ and 
$n(0) = 100$ particles. 
The results of a MCWF simulation averaged over 100 runs
are plotted as a thin solid line, while the mean-field results are 
plotted as a thick line.
}
\end{figure}

Parts (a) and (b) of the figure show the phase space structure without
dissipation and $Un = 0$ and $Un = 4J$, respectively. One observes the
familiar self-trapping bifurcation of the fixed points for $Un > 2J$
\cite{Smer97,Vard01}.
The phase space structure is significantly altered in the presence of
particle loss as shown in part (c). The most important consequence 
is the occurrence of an attractive and a repulsive fixed point instead 
of the elliptic fixed points in the dissipation-free case \cite{08mfdecay}.

In the course of the time the system will thus relax to the 
attractive stationary state illustrated Fig.~\ref{fig-spheres} (c).
A many-particle quantum state can now be represented by a 
quasi-distribution function on this classical phase space,
for instance the Husimi Q-function \cite{08phase1,08phase2}. 
In this picture, a pure BEC is represented by a maximally localized distribution 
function and the loss of purity corresponds to a broadening or
distortion of the Q-function. The existence of an attractive 
fixed point clearly leads to the contraction of a phase space 
distribution function and thus to a re-purification of the
many-particle quantum state as observed in Fig.~\ref{fig-strong-pa1} (a).

\begin{figure}[tb]
\centering
\includegraphics[width=8cm, angle=0]{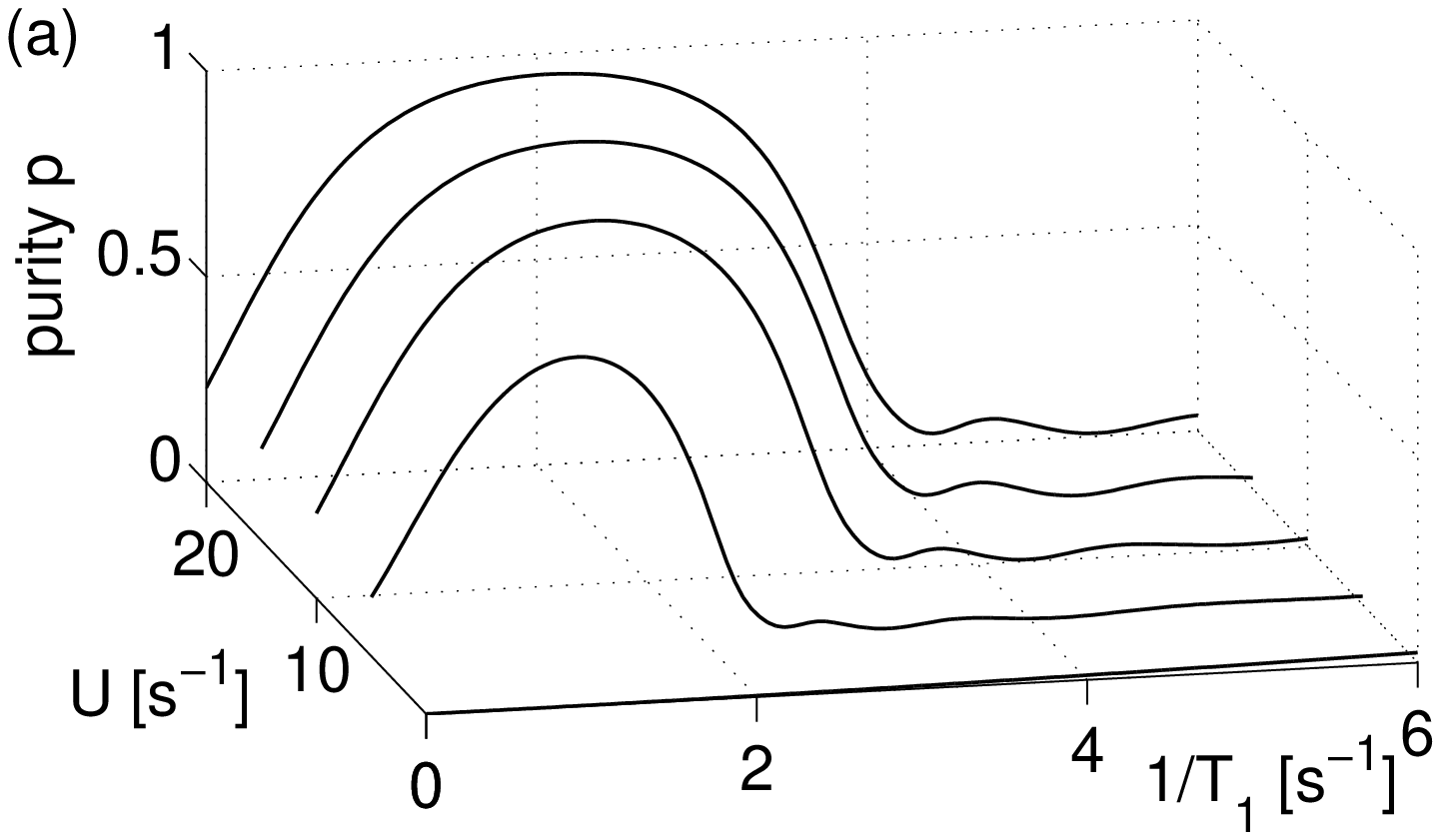}
\includegraphics[width=8cm, angle=0]{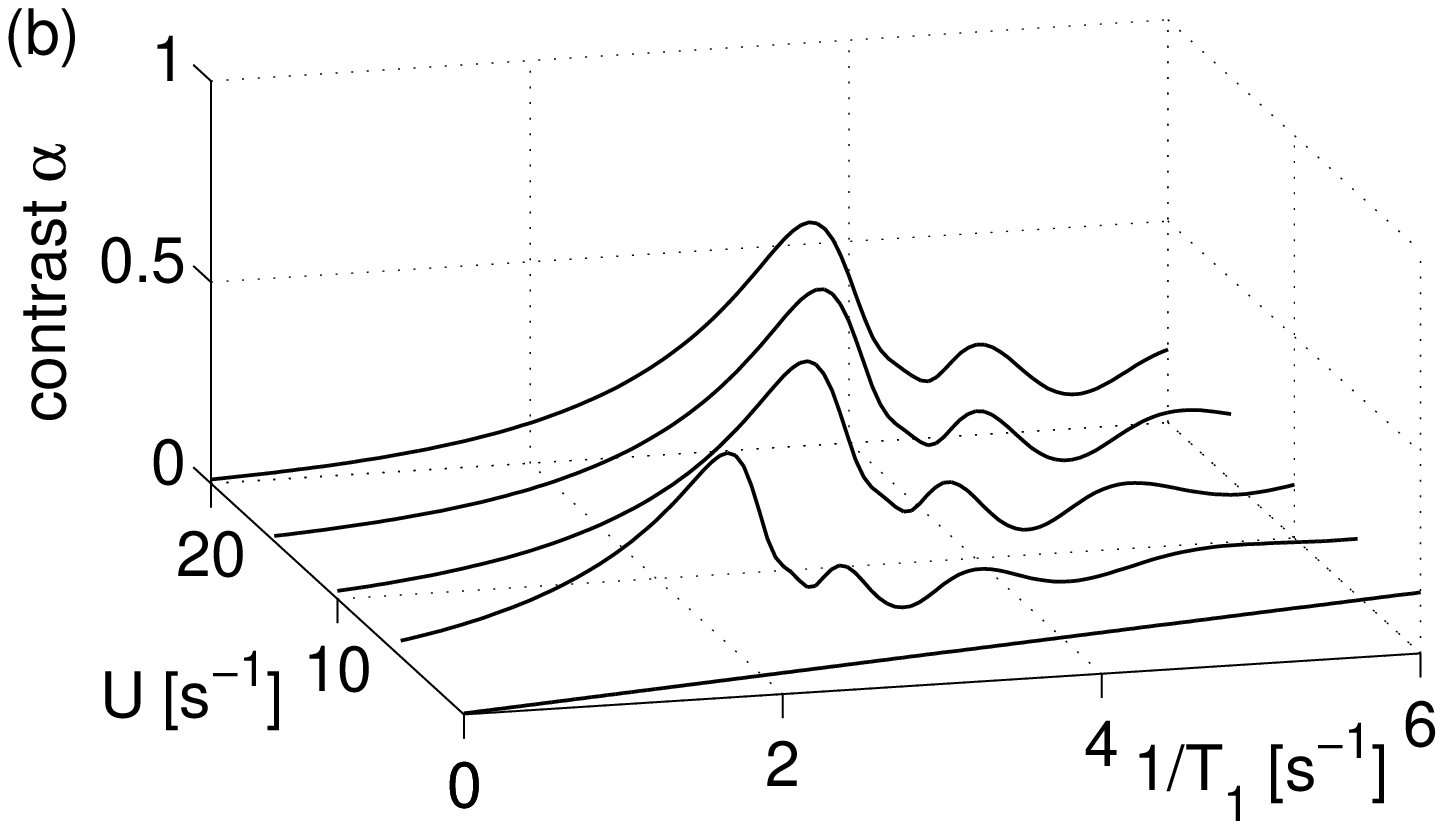}
\caption{\label{fig-coh-5s}
Purity $p$ (a) and contrast $\alpha$ (b) after $t = 2 \, {\rm s}$
as a function of the dissipation rate $1/T_1$ for different values
of the interaction strength $Un$  calculated within the mean-field 
approximation. The remaining parameters are chosen as in 
Fig.~\ref{fig-strong-pa1} (a).
}
\end{figure}

However, this nonlinear stationary state exists only
as long as the particle number exceeds a critical value given
by (cf. \cite{08mfdecay})
\be
  U^2 n^2 \apprge 4J^2 - f_a^2 T_1^{-2} \, .
  \label{eqn-critU}
\ee
As particles are slowly lost from the trap, the particle number eventually
falls below the critical value. For this reason the attractive fixed point
vanishes and the purity drops to the values expected for the linear 
case $U=0$. 
Since the attractive fixed point tends towards the equator maximizing $s_x/|\vec s|$,
the contrast assumes a maximum just before the disappearance of the 
attractive fixed point, while the purity is still large. 
In Fig.~\ref{fig-strong-pa1} (a) this happens after approx. 2.5 s.

The surprising effect of the re-purification of a BEC is extremely
robust -- it is present as long as the condition (\ref{eqn-critU})
is satisfied. A variation of the system parameters does not destroy
or significantly weaken the effect, it only changes the time scales
of this relaxation process. Fig.~\ref{fig-strong-pa2} compares the
time evolution of the purity and the contrast for three different
values of the particle loss rate $T_1^{-1}$. With increasing losses,
the nonlinear stationary state is reached much faster, but is also
lost earlier. 
One can thus maximize the purity or the contrast at a given point 
of time by engineering the loss rate. This effect is further illustrated in
Fig.~\ref{fig-coh-5s}, where the purity and the contrast after
2 seconds of propagation are shown in dependence of the loss rate
$T_1^{-1}$. Both curves assume a maximum for a certain finite value 
of $T_1^{-1}$.

\section{Conclusion and Outlook}

In summary, we have shown that the coherence properties of a weakly
and, in particular, also of a strongly interacting 
Bose-Einstein condensate in a double-well trap can be controlled 
by engineering the system's parameters and dissipation simultaneously. 
Surprisingly, dissipation can be used to stimulate coherence in the 
system, rather than -- as may be expected -- solely reduce it.

In the weakly interacting case, the contrast of the quasi-steady state 
of the system assumes a maximum for a finite value of the tunneling and the
dissipation rate. This stochastic resonance effect is robust against 
parameter variations. 
A Monte Carlo wave function simulation of the full many-body quantum dynamics
shows a good agreement to the mean-field description and provides a microscopic 
explanation of the observed effect. Moreover, a similar effect can be observed 
in the case where either the tunneling or the energy bias is driven, which is 
conceptually even closer to the common interpretation of stochastic resonance. 

In the last section, we have studied the effects of dissipation on the strongly 
interacting system. An important conclusion is that the interplay of interactions and 
dissipation can drive the system to a state of maximum coherence, while both 
processes alone usually lead to a loss of coherence.
We show that this effect can be understood from the appearance of an attractive 
fixed point in the mean-field dynamics reflecting the metastable behaviour of
the many-particle system.  

Since the double-well BEC is nowadays routinely realized with nearly 
perfect control on atom-atom interactions and external potentials 
\cite{Albi05,Gati06a,Gati06b}, we hope for an experimental verification 
of the predicted stochastic resonance effect.
An interesting perspective is to lift our results to extended 
dissipative setups, as e.g. studied in \cite{Hoff07,Hoff08}.
Besides the general idea of controlling many-body dynamics \cite{Madr06},
one may also investigate the possibility of dynamically engineering
entanglement in similar systems, as to some extend possible in 
state-of-the-art experiments \cite{Este08}.

\begin{acknowledgments}

We thank M.~K.~Oberthaler, J.~R.~Anglin and A.~S.~S\o{}rensen
for stimulating discussions.
This work has been supported by the German Research Foundation (DFG) 
through the research fellowship program (grant number WI 3415/1) and 
the Heidelberg Graduate School of Fundamental Physics (grant number 
GSC 129/1) as well as the Studienstiftung des deutschen Volkes.

\end{acknowledgments}



\begin{thebibliography}{0}
\expandafter\ifx\csname natexlab\endcsname\relax\def\natexlab#1{#1}\fi
\expandafter\ifx\csname bibnamefont\endcsname\relax
  \def\bibnamefont#1{#1}\fi
\expandafter\ifx\csname bibfnamefont\endcsname\relax
  \def\bibfnamefont#1{#1}\fi
\expandafter\ifx\csname citenamefont\endcsname\relax
  \def\citenamefont#1{#1}\fi
\expandafter\ifx\csname url\endcsname\relax
  \def\url#1{\texttt{#1}}\fi
\expandafter\ifx\csname urlprefix\endcsname\relax\def\urlprefix{URL }\fi
\providecommand{\bibinfo}[2]{#2}
\providecommand{\eprint}[2][]{\url{#2}}

\end{thebibliography}


\begin{thebibliography}{10}


\bibitem{Benz81} 
R.~Benzi, A.~Sutera, and A.~Vulpiani, 
J. Phys. A: Math. Gen. {\bf 14}, L453 (1981).

\bibitem{Wies95} 
K.~Wiesenfeld and F.~Moss, 
Nature {\bf 373}, 33 (1995).

\bibitem{Dykm95}
M.~I.~Dykman, D.~G.~Luchinsky, R.~Mannella, P.~V.~E.~McClintock, 
N.~D.~Stein, and N.~G.~Stocks, 
Nuovo Cimento D {\bf 17}, 661 (1995).

\bibitem{Gamm98} 
L.~Gammaitoni, P.~H\"anggi, P.~Jung, and F.~Marchesoni,
Rev. Mod. Phys. {\bf 70}, 223 (1998).

\bibitem{Well04}
T.~Wellens, V.~Shatokhin, and A.~Buchleitner,
Rep. Prog. Phys. {\bf 67}, 45 (2004).

\bibitem{Lofs94a}
R.~L\"{o}fstedt and S.~N.~Coppersmith, 
Phys. Rev. Lett. {\bf 72}, 1947 (1994).

\bibitem{Lofs94b}
R.~L\"{o}fstedt and S.~N.~Coppersmith, 
Phys. Rev. E {\bf 49}, 4821 (1994).

\bibitem{Buch98}
A.~Buchleitner and R.~N.~Mantegna
Phys. Rev. Lett. {\bf 80}, 3932 (1998).

\bibitem{Well99}
T.~Wellens and A.~Buchleitner, 
J. Phys. A {\bf 32}, 2895 (1999).

\bibitem{Huel00}
S.~F.~Huelga and M.~B.~Plenio, 
Phys. Rev. A {\bf 62}, 052111 (2000).

\bibitem{Adam01}
H.~H.~Adamyan, S.~B.~Manvelyan, and G.~Y.~Kryuchkyan, 
Phys. Rev. A {\bf 63}, 022102 (2001).

\bibitem{Huel07}
S.~F.~Huelga and M.~B.~Plenio, 
Phys. Rev. Lett. {\bf 98}, 170601 (2007).

\bibitem{Khod08}
Y.~Khodorkovsky, G.~Kurizki, and A.~Vardi,
Phys. Rev. Lett. {\bf 100}, 220403 (2008).

\bibitem{Li08}
Yun Li, Y.~Castin, and A.~Sinatra,
Phys. Rev. Lett. {\bf 100}, 210401 (2008).

\bibitem{Krau08}
B.~Kraus, H.~P.~B\"{u}chler, S.~Diehl, A.~Kantian, A.~Micheli, and P.~Zoller
Phys. Rev. A {\bf 78}, 042307 (2008).

\bibitem{Diel08}
S.~Diehl, A.~Micheli, A.~Kantian, B.~Kraus, H.~P.~B\"{u}chler, and P.~Zoller,
Nature Physics (in print), arXiv:0803.1482v1.

\bibitem{Vers08}
F.~Verstraete, M.~M.~Wolf, and J.~I.~Cirac, 
Nature Physics (accepted), arXiv:0803.1447v2.

\bibitem{Syas08}
N.~Syassen, D.~M.~Bauer, M.~Lettner, T.~Volz, D.~Dietze, J.~J.~Garcia-Ripoll,
J.~I.~Cirac, G.~Rempe, and S.~D\"{u}rr,
Science {\bf 320}, 1329 (2008).

\bibitem{Garc08}
J.~J.~Garcia-Ripoll, S.~D\"{u}rr, N.~Syassen, D.~M.~Bauer, M.~Lettner, G.~Rempe, 
and J.~I.~Cirac, arXiv:0809.3679v1.

\bibitem{08stores}
D.~Witthaut, F.~Trimborn, and S.~Wimberger,
Phys.~Rev.~Lett. (in print), preprint arXiv:0809.1776v1 (2008).

\bibitem{Albi05}
M.~Albiez, R.~Gati, J.~F\"olling, S.~Hunsmann, M.~Cristiani, and M.~K.~Oberthaler,  
Phys. Rev. Lett.  {\bf 95}, 010402  (2005).

\bibitem{Gati06a}
R.~Gati, B.~Hemmerling, J.~F{\"o}lling, M.~Albiez, and M.~K.~Oberthaler,  
Phys. Rev. Lett.  {\bf 96}, 130404  (2006).

\bibitem{Gati06b}
R.~Gati, J.~Est\`eve, B.~Hemmerling, T.B. Ottenstein, J.~Appmeier, A.~Weller, and M.~K.~Oberthaler,  
New J. Phys.  {\bf 8}, 189  (2006).

\bibitem{Foll07}
S.~F\"olling, S.~Trotzky, P.~Cheinet, M.~Feld, R.~Saers, 
A.~Widera, T.~M\"uller, and I.~Bloch, 
Nature {\bf 448}, 1029 (2007).

\bibitem{Trot08}
S.~Trotzky, P.~Cheinet, S.~F\"olling, M.~Feld, U.~Schnorrberger, A.~M.~Rey, 
A.~Polkovnikov, E.~A.~Demler, M.~D.~Lukin, and I.~Bloch,
Science {\bf 319},  295 (2008).

\bibitem{Schu05b}
T.~Schumm, S.~Hofferberth, L.~M. Andersson, S.~Wildermuth, S.~Groth,
  I.~Bar-Joseph, J.~Schmiedmayer, and P.~Kr\"{u}ger,  
  Nature Physics  {\bf 1}, 57  (2005).

\bibitem{Milb97}
G.~J. Milburn, J.~Corney, E.~M. Wright, and D.~F. Walls,  
Phys. Rev. A  {\bf 55}, 4318  (1997).

\bibitem{Smer97}
A.~Smerzi, S.~Fantoni, S.~Giovanazzi, and S.~R. Shenoy,  
Phys. Rev. Lett.  {\bf 79}, 4950  (1997).

\bibitem{Vard01}
A.~Vardi and J.~R. Anglin,  
Phys. Rev. Lett.  {\bf 86}, 568  (2001).

\bibitem{Angl01}
J.~R. Anglin and A.~Vardi,  
Phys. Rev. A  {\bf 64}, 013605  (2001).

\bibitem{08phase1}
F.~Trimborn, D.~Witthaut, and H.~J.~Korsch, 
Phys. Rev. A {\bf 77}, 043631 (2008).

\bibitem{08phase2}
F.~Trimborn, D.~Witthaut, and H.~J.~Korsch, 
arXiv:0802.1142v2.

\bibitem{Angl97}
J.~R.~Anglin,  Phys. Rev. Lett.  {\bf 79}, 6  (1997).

\bibitem{Ruos98}
J.~Ruostekoski and D.~F.~Walls,  
Phys. Rev. A  {\bf 58}, R50  (1998).

\bibitem{Wall85}
D.~F. Walls and G.~J. Milburn,  
Phys. Rev. A  {\bf 31}, 2403  (1985).

\bibitem{Gard04}
C.~W.~Gardiner and P.~Zoller, {\it Quantum Noise}, 
Springer Series in Synergetics, Berlin (2004).

\bibitem{Bloc99}
I.~Bloch, T.~W.~H\"{a}nsch, and T.~Esslinger,
Phys. Rev. Lett. {\bf 82}, 3008  (1999).

\bibitem{08mfdecay}
F.~Trimborn, D.~Witthaut, and S.~Wimberger,
J.~Phys.~B:~At.~Mol.~Opt.~Phys. {\bf 41}, 171001 (FTC) (2008).

\bibitem{Viol00}
L.~Viola, E.~M.~Fortunato, S.~Lloyd, C.-H.~Tseng, and D.~G.~Cory,
Phys. Rev. Lett. {\bf 84}, 5466  (2000).

\bibitem{Bloc46}
F.~Bloch, 
Phys. Rev. {\bf 70}, 460 (1946).

\bibitem{Legg01}
A.~J. Leggett,  Rev. Mod. Phys.  {\bf 73}, 307  (2001).

\bibitem{Wimb05}
S.~Wimberger, R.~Mannella, O.~Morsch, E.~Arimondo, A.~R.~Kolovsky, and A.~Buchleitner,
Phys. Rev. A {\bf 72}, 063610 (2005).

\bibitem{Schl06}
P. Schlagheck and T. Paul, Phys. Rev. A {\bf 73}, 023619 (2006).

\bibitem{Schl07}
P. Schlagheck and S. Wimberger, Appl. Phys. B {\bf 86}, 385 (2007).

\bibitem{Wu00}
Biao Wu and Qian Niu, Phys. Rev. A {\bf 61}, 023402 (2000).

\bibitem{Wu06}
Biao Wu and Jie Liu, Phys. Rev. Lett. {\bf 96}, 020405 (2006).

\bibitem{06LZnonlin1}
D.~Witthaut, E.~M.~Graefe, and H.~J.~Korsch, Phys. Rev. A {\bf 73}, 063609 (2006).

\bibitem{06LZnonlin2}
E.~M.~Graefe, H.~J.~Korsch, and D.~Witthaut, Phys. Rev. A {\bf 73}, 013617 (2006).

\bibitem{Paul05}
T.~Paul , K.~Richter, and P.~Schlagheck,
Phys. Rev. Lett. {\bf 94}, 020404 (2005).

\bibitem{06nl_transport}
K.~Rapedius, D.~Witthaut, and H.~J.~Korsch,
Phys. Rev. A {\bf 73}, 033608 (2006).  

\bibitem{Dali92}
J.~Dalibard, Y.~Castin, and K.~M\o{}lmer, 
Phys. Rev. Lett. {\bf 68}, 580 (1992).

\bibitem{Molm93}
K.~M\o{}lmer, Y.~Castin, and J.~Dalibard,
J. Opt. Soc. Am. B {\bf 10}, 524 (1993).

\bibitem{Carm93}
H.~J.~Carmichael, \textit{An Open Systems Approach to Quantum Optics}, 
Springer, Berlin (1993).

\bibitem{Cris04}
M.~Cristiani, O.~Morsch, N.~Malossi, M.~Jona-Lasinio, M.~Anderlini, 
E.~Courtade, and E.~Arimondo,
Opt. Express {\bf 12}, 4-10 (2004).

\bibitem{Hoff07}
S.~Hofferberth, I.~Lesanovsky, B.~Fischer, T.~Schumm, and J.~Schmiedmayer,
Nature {\bf 449}, 324 (2007).

\bibitem{Hoff08}
S.~Hofferberth, I.~Lesanovsky, T.~Schumm, J.~Schmiedmayer, A.~Imambekov, V.~Gritsev, and E.~Demler,
Nature Physics {\bf 4}, 489 (2008).

\bibitem{Madr06}
J.~Madro\~{n}ero, A.~Ponomarev, A.~R.~R.~Carvalho, S.~Wimberger, C.~Viviescas, A.~Kolovsky, 
K.~Hornberger, P.~Schlagheck, A.~Krug, and A.~Buchleitner,
Adv. At. Mol. Opt. Phys. {\bf 53}, 33 (2006).

\bibitem{Este08}
J.~Est\`eve, C.~Gross, A.~Weller, S.~Giovanazzi, and M.~K.~Oberthaler,
Nature (in print), arXiv:0810.0600v1.

\end{thebibliography}
\end{document}